\documentclass[lettersize,journal]{IEEEtran}
\usepackage{amsmath,amsfonts}
\usepackage{algorithmic}
\usepackage{array}
\usepackage{textcomp}
\usepackage{stfloats}
\usepackage{url}
\usepackage{verbatim}
\usepackage{graphicx}

\usepackage{mathtools, amssymb}
\usepackage{braket}
\usepackage{subcaption}
\newcommand{\Conv}{%
  \mathop{\scalebox{1.5}{\raisebox{-0.2ex}{$\circledast$}}
  }
}

\def\BibTeX{{\rm B\kern-.05em{\sc i\kern-.025em b}\kern-.08em
    T\kern-.1667em\lower.7ex\hbox{E}\kern-.125emX}}
\usepackage{balance}
\begin{document}
\title{Myopic Entropy Scheduling for Ramsey
Magnetometry}
\author{\uppercase{Julian Greentree}\iffalse\authorrefmark{1}\fi,  
\uppercase{William Moran}% \IEEEmembership{Life Member}
, Robin Evans% \IEEEmembership{Life Fellow}
, Andrew Melatos, Neel Kanth Kundu,% \IEEEmembership{Member}
 and Peter M. Farrell
\thanks{J. Greentree, W. Moran, R. Evans and G. Nair are with the Faculty of Engineering and Information Technology, University of Melbourne, Melbourne, Australia} \thanks{A. Melatos is with the School of Physics, University of Melbourne, Melbourne, Australia.}\thanks{N. K. Kundu is with the Centre for Applied Research in Electronics, Indian Institute of Technology Delhi, New Delhi, India}\thanks{P.M. Farrel is with MOGLabs, Melbourne,  Australia} \thanks{J. Greentree is supported by an Australian Government Research Training Program (RTP) Scholarship.}}

\markboth
{Greentree \emph{et al.}: Myopic Entropy Scheduling for Ramsey
Magnetometry}
{Greentree \emph{et al.}: Myopic Entropy Scheduling for Ramsey
Magnetometry}

\iffalse{\corresp{Corresponding author: Julian Greentree (e-mail: jgreentree@unimelb.edu.au).}}\fi
\maketitle

\begin{abstract}
This paper presents an entropy based adaptive measurement sequence strategy for quantum sensing of magnetic fields. To physically ground our ideas we consider a sensor employing a nitrogen vacancy center in diamond, however our approach is applicable to other quantum sensor arrangements. The sensitivity and accuracy of these sensors typically rely on long sequences of rapidly occurring measurements. We introduce a new  technique for designing these measurement sequences aimed at reducing the number of measurements required for a specified accuracy as measured by entropy, by selecting measurement parameters that optimally reduce entropy at each measurement. We compare, via simulation, the efficiency and sensitivity of our new method with several existing measurement sequence design strategies. Our results show quantifiable improvements in sensing performance. We also show analytically that our entropy reduction approach, reduces, under certain simplified conditions, to a well-known and widely used measurement strategy.
\end{abstract}

\begin{IEEEkeywords}
Bayesian modelling, Entropy-based sensing, Nitrogen Vacancy Magnetometry, Quantum information, Sensor Scheduling, Shannon Entropy.
\end{IEEEkeywords}

\section{Introduction}
\label{sec:introduction}

\IEEEPARstart{L}{aser-based} quantum sensing often relies on a sequence of timed and tailored pulses acting on a controlled quantum state, generating a corresponding sequence of readout information from which physical quantities can be inferred \cite{bonatoOptimizedQuantumSensing2016, degenQuantumSensing2017, barrySensitivityOptimizationNVdiamond2020, woodLongSpinCoherence2022, higginsEntanglementfreeHeisenberglimitedPhase2007, higginsDemonstratingHeisenberglimitedUnambiguous2009, shiga_locking_2012}. In this work, we investigate the adaptive properties of one such experiment, the Ramsey experiment, for nitrogen-vacancy (NV) magnetometry, and the use of an entropy minimisation approach for adaptive sensing.

The NV centre is a point defect in a diamond lattice, a nitrogen atom and adjacent vacancy in the place of two carbon atoms. The NV centre, when negatively charged (often written NV\textsuperscript{-}) contains a bound electron that is sensitive to magnetic fields\cite{barrySensitivityOptimizationNVdiamond2020}. When an external bias field $B_0$ is applied, collinear with the NV centre, the mechanics can be well modeled by a two level system with energy levels determined by Zeeman splitting\cite{bonatoOptimizedQuantumSensing2016, barrySensitivityOptimizationNVdiamond2020}. The ground state $\ket{0}$, corresponds to the electron's magnetic moment, $\mu$ (otherwise defined in\iffalse Eq.\fi~\eqref{def: nv hamiltonian}, $\approx2.8\times10^{10}\text{Hz}/\text{T}$ \cite{barrySensitivityOptimizationNVdiamond2020}) aligning to the external field, and the excited state $\ket{1}$ anti-aligned, thus, when the system is exposed to an aligned magnetic field of unknown magnitude $B$, the Hamiltonian, can be written as:
\begin{align}
    \hat{H} = -\frac{\hbar\mu}{2}(B+B_0) (\ket{0}\bra{0} - \ket{1}\bra{1}).\label{def: nv hamiltonian}
\end{align}
Many measurement protocols have been devised to make use of this phenomenon \cite{barrySensitivityOptimizationNVdiamond2020, santagatiMagneticFieldLearningUsing2019}. Here we focus on one widely used method, the Ramsey experiment, where the NV\textsuperscript{-} system is excited by a $\pi/2$ laser pulse, tuned to the energy difference between the ground state and the excited state. This pulse excites the system into an equal superposition of the ground and excited state, from which it undergoes Larmor precession with a rate proportional to the external magnetic field \cite{santagatiMagneticFieldLearningUsing2019, woodLongSpinCoherence2022} for some time $\tau$. To read the signal, a second laser pulse is applied to the state, with a phase $\theta$. This read out pulse converts the system into either the excited state, which will lead to a photon emission event or the ground state, which does not. The detection of this photon is a non-trivial undertaking, and mulitple read-out methods exist \cite{degenQuantumSensing2017, barrySensitivityOptimizationNVdiamond2020}. Detection difficulties of this form can be modelled by a rescaling of the likelihood function reducing the contrast of the signal or introducing a bias in favor of detection or non-detection. In this work we focus on the quantum mechanism directly, where perfect detection is assumed. In the absence of photon detection concerns, the likelihood of emission/nonemission events can be expressed\cite{degenQuantumSensing2017, craigieResourceefficientAdaptiveBayesian2021}:
\begin{align}
    \Pr(X = 0|B,\tau,\theta) &= 1/2+e^{-\tau/T}\cos(2\mu B\tau + \theta)/2\label{def: pr(x=0) likelihood function}\\
    \Pr(X = 1|B,\tau,\theta) &= 1/2-e^{-\tau/T}\cos(2\mu B\tau + \theta)/2,\label{def: pr(x=1) likelihood function}
\end{align}
where $\tau$ is the exposure time, $\theta$ is the phase readout parameter (encoded in the relative phase difference between the first and second readout pulse), and $T$ is the coherence time of the state, incorporating dephasing noise \cite{degenQuantumSensing2017, barrySensitivityOptimizationNVdiamond2020, santagatiMagneticFieldLearningUsing2019, craigieResourceefficientAdaptiveBayesian2021}. This amounts to sinusoidal oscillation between peaks as the state precesses smoothly, with decaying contrast as the state is overwhelmed by external noise encoded within the $e^{-\tau/T}$ coefficient. In the form shown in \iffalse \iffalse Eq.~\fi\fi\eqref{def: pr(x=0) likelihood function} and \eqref{def: pr(x=1) likelihood function} the Ramsey experiment provides two tunable measurement parameters $\theta$, $\tau$. 

There are multiple methods for parameter selections that have been designed and implemented, for example, so called linear sensing methods focus on maximising the slope in the likelihood function  and calculating the most likely magnetic field value from the resultant probabilities. In this approach sensitivity is limited by the coherence time $T$ and number of individual Ramsey measurements conducted \cite{degenQuantumSensing2017}. \iffalse{Optimal parameters locate the low bias/high slope region of the likelihood function over the majority of the prior and chooses $\tau$ to maximise $\partial/\partial B \Pr(X=x_i|B, \tau, \theta)$, }\fi The posterior variance decreases linearly with $T$ and with the square root of the number of measurements $N_\text{count}$ \cite{degenQuantumSensing2017, craigieResourceefficientAdaptiveBayesian2021}. Sensitivity scaling can also be framed in terms of a ``quantum resource," total exposure time $N$ (noting that $N=\tau N_\text{count}$)\cite{degenQuantumSensing2017,higginsEntanglementfreeHeisenberglimitedPhase2007}. This $\sqrt{N}$ scaling defines the ``standard quantum limit". An improved Heisenberg limited (HL) scaling has been demonstrated, yielding a linear scaling in $N$ \cite{degenQuantumSensing2017, higginsEntanglementfreeHeisenberglimitedPhase2007,higginsDemonstratingHeisenberglimitedUnambiguous2009, kitaevQuantumMeasurementsAbelian1996, berryHowPerformMost2009, ahmadiQuantumPhaseEstimation2012}. It can also be framed as an entropic relationship\iffalse, noting that $H(\tau)+H(B)\leq C$\fi  \cite{maassenGeneralizedEntropicUncertainty1988,ban_state_1999,mehrmanesh_effects_2024}. HL scaling can be achieved by entangled quantum measurements but also by dynamic or adaptive sensing regimes \cite{bonatoOptimizedQuantumSensing2016,degenQuantumSensing2017, barrySensitivityOptimizationNVdiamond2020, craigieResourceefficientAdaptiveBayesian2021, higginsEntanglementfreeHeisenberglimitedPhase2007, higginsDemonstratingHeisenberglimitedUnambiguous2009, kitaevQuantumMeasurementsAbelian1996, ahmadiQuantumPhaseEstimation2012, giovannettiQuantumEnhancedMeasurementsBeating2004}. An adaptive measurement scheme can be designed in many ways. There is a strong analogy that can be drawn with phase estimation in an interferometric, quantum computing and solid state sensing contexts \cite{degenQuantumSensing2017, higginsEntanglementfreeHeisenberglimitedPhase2007}. This can be justified from a physical perspective, noting the similarity between the phase encoding of the magnetic field in the NV centre and the phase measurement problem; or from a mathematical perspective, noting the similarity in form of the relevant probability distributions. An early measurement scheme that leverages quantum measurements in a computation context was proposed by Kitaev \cite{kitaevQuantumMeasurementsAbelian1996} and adapted to an adaptive algorithm that can be applied in an interferometric context \cite{higginsEntanglementfreeHeisenberglimitedPhase2007} and implimented in an NV context \cite{bonatoOptimizedQuantumSensing2016}. This algorithm achieves HL scaling in the absence of decoherence ($T\rightarrow\infty$) \cite{kitaevQuantumMeasurementsAbelian1996, higginsEntanglementfreeHeisenberglimitedPhase2007}, and can be adapted to allow for $1/N$ scaling in the presence of decoherence, though at the cost of a reduction in variance decrease by a scale factor \cite{higginsEntanglementfreeHeisenberglimitedPhase2007, higginsDemonstratingHeisenberglimitedUnambiguous2009}.

\section{Problem statement}
To make explicit the problem, we are trying to measure a static magnetic field $B$, using a sequence of adaptive Ramsey measurements. We seek to construct a policy to select measurement parameters ($\tau$ and $\theta$) that will reduce the posterior entropy of the magnetic field distribution after Bayesian updating: $\Pr(B|\{X_i=x_i, \theta_i, \tau_i\})$, for some adaptive sequence of measurements $\{X_i\}$.

\subsection{Mathematical notation}
\iffalse In this work, we discuss probability distributions written $\Pr(X=x)$ where $X$ is either a discrete or continuous variable, taking the value $x$. In the specific case of the magnetic field strength, $B$, the variable and its value both share the same symbol and thus $\Pr(B=B)$ is shortened to $\Pr(B)$. The conditional probability \fi

When describing a policy, we need a map from the state space of our system to a future measurement. In the NV case, this state is described entirely by the posterior distribution, or equivalently the sequence of prior measurement parameters and outcomes $\{\tau_i, \theta_i, x_i\}$.
There are many methods to generate maps from the state to measurement parameters, resulting in different adaptive protocols \cite{degenQuantumSensing2017}, Bayesian posterior calculation allows for optimisation of some functional over possible measurement outcomes. Key functionals of note are the variance ($V[\Pr(B)]=\left\langle B^2\right\rangle-\left\langle B\right\rangle^2$) and Shannon entropy ($H[\Pr(B)]=-\int \Pr(B)\ln(\Pr(B))dB$) \cite{nielsenQuantumComputationQuantum2010, coverThomasElementsofInformationTheory}. For any given criterion of minimisation $g$, we can define the functional $G$ mapping from non-negative, $L_1$ normalised functions to $\mathbb{R}$ (such that $G(\Pr(B)) = g(\Pr(B))$), and define optimal measurement parameters ($\tau_\text{min}, \theta_\text{min}$) by minimising the expected value of $G$ given some measurement $X$:

\begin{align}
    \left\langle G|X\right\rangle &= \sum_{X=x_i}\Pr(X=x_i)G[\Pr(B|X=x_i)]\\
    \tau_\text{min}, \theta_\text{min} &\coloneqq\text{argmin}_{\tau,\theta} \left\langle G|X\right\rangle.
\end{align}
In the context of entropy minimisation, this is equivalent to maximising mutual information \cite{nielsenQuantumComputationQuantum2010}:
\begin{align}
    \left\langle H|X\right\rangle = H(B|X) &= H(B) - I(B;X)
\end{align}
Since the entropy of the prior, $H(B)$, is independent of any measurement parameters:
\begin{align}
    \text{argmin}_{\tau,\theta} \left\langle H|X\right\rangle &= \text{argmax}_{\tau,\theta} I(B;X_{\tau,\theta})\\
    &=\text{argmax}_{\tau,\theta} H(X_{\tau,\theta})-H(X_{\tau,\theta}|B).
\end{align}
The mutual information, $I$, can be expressed directly from the prior distribution and likelihood functions \cite{coverThomasElementsofInformationTheory}:
\begin{align}
    I(B;X_{\tau,\theta}) &= -\sum_{X=x_i} \left[\left(\int\Pr(X=x_i|B,\tau,\theta)\Pr(B)dB\right)\right.\nonumber\\
    &\quad\left.\times\ln\left(\int\Pr(X=x_i|B,\tau,\theta)\Pr(B)dB\right)\right]\nonumber\\
    &\quad+\int \sum_{X=x_i}\left[\Pr(X=x_i|B,\tau,\theta)\Pr(B)\right.\nonumber\\
    &\quad\left.\times\ln\left(\Pr(X=x_i|B,\tau,\theta)\right)\right]\Pr(B)dB.\label{eq: mutual information from the prior}
\end{align}
A simple derivation of this equality is provided in Appendix \ref{app: Mutual information derivation}. 

Choosing measurement parameters to maximise mutual information constitutes a myopic approach to minimising posterior entropy. A more complex method could be considered wherein the total mutual information is maximised over a string of measurements. Such an approach would be significantly more algorithmically intensive and due to the submodularity of the entropy minimisation problem, would deliver at most a constant increase in performance (of less that 60\%) \cite{addabboDynamicalSystemsBasedHierarchy2019, iyer_generalized_2022,wolseyAnalysisGreedyAlgorithm1982}.Compatibility with the myopic approximation distinguishes entropy reduction from other approaches (\emph{e.g.}, variance reduction) which, while capable of similar performances cannot guarantee optimal scaling without considering future measurements.

\subsection{Connections to previous algorithms}

The optimal measurement question has been tackled previously, both in the context of NV magnetometry \cite{bonatoOptimizedQuantumSensing2016, degenQuantumSensing2017, barrySensitivityOptimizationNVdiamond2020, santagatiMagneticFieldLearningUsing2019, craigieResourceefficientAdaptiveBayesian2021,berryHowPerformMost2009} and in the context of interferometry which possesses analygous likelihood functions \cite{higginsEntanglementfreeHeisenberglimitedPhase2007, higginsDemonstratingHeisenberglimitedUnambiguous2009, kitaevQuantumMeasurementsAbelian1996, berryHowPerformMost2009, ahmadiQuantumPhaseEstimation2012, svoreFasterPhaseEstimation2014}. In the absence of decoherence (decoherence time $T\rightarrow\infty$) an optimal measurement policy was presented by Kitaev \cite{kitaevQuantumMeasurementsAbelian1996}, which is optimal; both in terms of ``quantum resources" (either total exposure time $\sum_i \tau_i$ in the context of NV magnetometry, or photon/phaseplate interaction in the case of interferometry) \cite{barrySensitivityOptimizationNVdiamond2020, kitaevQuantumMeasurementsAbelian1996, berryHowPerformMost2009}, and in terms of measurement number as we show in this work. The algorithm presented in Ref.~\cite{kitaevQuantumMeasurementsAbelian1996} has been re-expressed in the form of an adaptive algorithm requiring no long range inter-measurement entanglement \cite{higginsEntanglementfreeHeisenberglimitedPhase2007, higginsDemonstratingHeisenberglimitedUnambiguous2009}. This allows for application in the NV magnetometry context without the need for coherent quantum control. In Sec.~\ref{sec: simplified case} we present a novel derivation of this well known optimal algorithm, using fundamental principles from information theory. Due to the broad applicability of the Bayesian probability framework and Shannon information, the approach presented here allows for measurement scheduling outside of the regime for which the algorithm first presented in Ref.~\cite{kitaevQuantumMeasurementsAbelian1996} is optimal.

\section{Bayesian Analysis}\label{sec: bayesian analysis}

\begin{figure}
    \centering
    \includegraphics[width=0.8\linewidth]{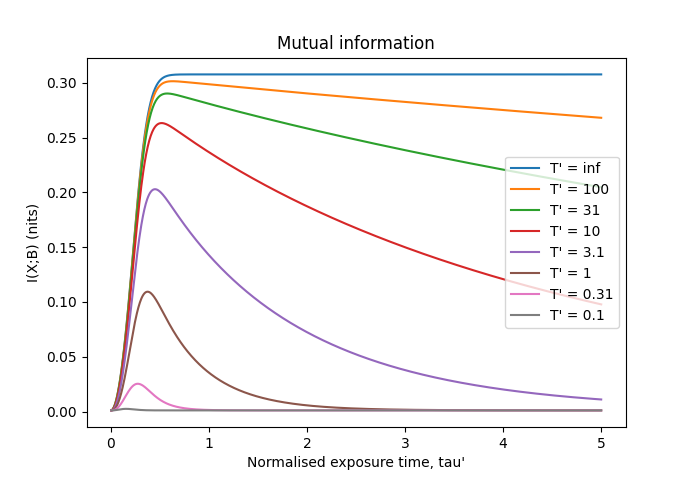}
    \caption{The mutual information between the measurement $X$ and magnetic field $B$, $I(X:B)$ over a Gaussian prior in $B$. All variables are quoted in natural units ($\tau', T' = \tau/\mu,\, T/\mu)$, $\theta$ is set to 0, and the standard deviation of the prior is set at $3/(\sqrt{2})$.}
    \label{fig: mutual information first measurement}
\end{figure}

Given the definition of the likelihood function given in \iffalse Eqs.\fi~\eqref{def: pr(x=0) likelihood function} and \eqref{def: pr(x=1) likelihood function} we can write \iffalse \iffalse Eq.~\fi\fi\eqref{eq: mutual information from the prior} for a Ramsey measurement on an arbitrary prior:
\begin{align}
    I(B;X_{\tau,\theta}) &= -\sum \left(\int\frac{1\pm e^{-\tau/T}\cos(2\mu B\tau + \theta)}{2}\Pr(B)dB\right)\nonumber\\
    &\quad\times\ln\left(\int\frac{1\pm e^{-\tau/T}\cos(2\mu B\tau + \theta)}{2}\Pr(B)dB\right)\nonumber\\
    &\quad+\int \sum\frac{1\pm e^{-\tau/T}\cos(2\mu B\tau + \theta)}{2}\nonumber\\
    &\quad\times\ln\left(\frac{1\pm e^{-\tau/T}\cos(2\mu B\tau + \theta)}{2}\right)\Pr(B)dB.\label{eq: mutual information for ramsey likelihoods}
\end{align}
This functional can be computed for any distribution $\Pr(B)$, phase ($\theta\in[0,2\pi)$ and exposure time ($\tau\geq0$). The value of the mutual information of a Ramsey measurement over a Gaussian prior is provided in Fig.~\ref{fig: mutual information first measurement} for varying values of $T$. For low exposure times, the localisation of the prior creates high bias, lowering $H(B)$ and thus limiting the mutual information. As the exposure time, $\tau$, increases, mutual information increases to a maximum before decoherence overwhelms the system, increasing conditional entropy, $H(X|B)$, and reducing mutual information to zero. The location and magnitude of the maxima depends positively on the coherence time, $T$, mirroring results from constant parameter sensing discussed further in \cite{degenQuantumSensing2017}.

%\textcolor{red}{Include Figure 1 here first measurement mutual information}

\iffalse When applied a Gaussian prior in the absence of decoherence there is a monotonic increase in mutual information yield with $\tau$. This arises from a redundancy between localising information encoded within the Gaussian prior and information that would be gained in low $\tau$ measurements, which creates a bias in the outcome probabilities, lowering $H(X_{\tau,\theta})$. In the presence of decoherence ($T<\infty$) increases in $I$ arising from bias reduction are offset by a monotonic increase in measurement noise caused by the decoherence which results in an increase in $H(X_{\tau,\theta}|B)$.\fi

Posteriors arising from future measurements of the initial Gaussian prior can be calculated, either directly from the posterior (substituting $\Pr(B|X=x_i,\tau,\theta)$ into \iffalse \iffalse Eq.~\fi\fi\eqref{eq: mutual information for ramsey likelihoods}) or explicitly, noting that in accordance with Bayes rule \cite{coverThomasElementsofInformationTheory, jeffreysTheoryProbability1998}:
\begin{align}
    \Pr(B|X=x_i, \tau, \theta) &= \frac{\left(1\pm e^{-\tau/T}\cos(2\mu B\tau + \theta)\right)Pr(B)}{\int1\pm e^{-\tau/T}\cos(2\mu B\tau + \theta)\Pr(B)dB}.
\end{align}
Future mutual information can thus be computed as:

\begin{multline}
    I(B;X_{\tau,\theta}|\{X_i=x_i\}) =\\
     -\sum \left(\int\frac{1\pm e^{-\tau/T}\cos(2\mu B\tau + \theta)}{2}\right.\\
    \quad\left.\frac{\prod_i1+ e^{-\tau_i/T}\cos(2\mu B\tau_i + \theta_i+\pi x_i)\Pr(B)}{\int\prod_i1+ e^{-\tau_i/T}\cos(2\mu B\tau_i + \theta_i+\pi x_i)\Pr(B)dB}dB\right)\\
    \quad\times\ln\left(\int\frac{1\pm e^{-\tau/T}\cos(2\mu B\tau + \theta)}{2}\right.\\
    \quad\left.\frac{\prod_i1+ e^{-\tau_i/T}\cos(2\mu B\tau_i + \theta_i+\pi x_i)\Pr(B)}{\int\prod_i1+ e^{-\tau_i/T}\cos(2\mu B\tau_i + \theta_i+\pi x_i)\Pr(B)dB}dB\right)\\
    \quad+\int \sum\frac{1\pm e^{-\tau/T}\cos(2\mu B\tau + \theta)}{2}\\
    \quad\ln\left(\frac{1\pm e^{-\tau/T}\cos(2\mu B\tau + \theta)}{2}\right)\\
    \quad\times\frac{\prod_i1+ e^{-\tau_i/T}\cos(2\mu B\tau_i + \theta_i+\pi x_i)}{\int\prod_i1+ e^{-\tau_i/T}\cos(2\mu B\tau_i + \theta_i+\pi x_i)\Pr(B)dB}\Pr(B)dB\label{eq: mutual information for ramsey likelihoods n}.
\end{multline}
    
Equations \eqref{eq: mutual information for ramsey likelihoods n} and \eqref{eq: mutual information for ramsey likelihoods} apply to different distributions, $\Pr(B|\{X_i=x_i, \tau_i, \theta_i\})$ and $\Pr(B)$. These distributions differ by the inclusion of a factor $\Gamma$, defined by their ratio. We can analyse this factor directly, noting in the Fourier domain: %\textcolor{red}{Need to check the exact coefficients for correctness}
\begin{multline}
    \Gamma(B, \{x_i, \tau_i,\theta_i\}) \coloneqq \frac{\Pr(B|X=x_i,\tau_i,\theta_i)}{\Pr(X=x_i, \tau_i\theta_i)}\\=\frac{\prod_i1+ e^{-\tau_i/T}\cos(2\mu B\tau_i + \theta_i+\pi x_i)}{\int\prod_i1+ e^{-\tau_i/T}\cos(2\mu B\tau_i + \theta_i+\pi x_i)\Pr(B)dB}.
\end{multline}
The likelihood function of each individual measurement contains only three frequency components, one at $\xi=0$ and two at $\xi=\pm2\mu B\tau_i$, we can thus write:
\begin{multline}
    \mathcal{F}_B[\Gamma(B, \{x_i, \tau_i,\theta_i\})](\xi)=\\
    \Conv_i [2\delta(\xi)+e^{-\tau_i/T+i\left(\frac{\theta_i+\pi x_i}{2\mu B\tau}\right)}\delta(\xi+2\mu\tau_i)\cr+e^{-\tau_i/T-i\left(\frac{\theta_i+\pi x_i}{2\mu B\tau}\right)}\delta(\xi-2\mu\tau_i)](\xi)\cr\times\left(\Conv_i [2\delta(\xi)+e^{-\tau_i/T+i\left(\frac{\theta_i+\pi x_i}{2\mu B\tau}\right)}\delta(\xi+2\mu\tau_i)\right.\cr\left.+e^{-\tau_i/T-i\left(\frac{\theta_i+\pi x_i}{2\mu B\tau}\right)}\delta(\xi-2\mu\tau_i)]\circledast \mathcal{F}[\Pr(B)](0)\right)^{-1}.
\end{multline}
Where $\mathcal{F}[\Pr(B)](\xi)$ is the inverse Fourier-Stieltjes transform of $\Pr(B)$ in $B$-space and $[A(x)\circledast B(y)](z)$ is the convolution  between two measures. Here, in keeping with the Physics notation, $\delta(\xi-a)$ is the unit point mass at $a$. We can gather each term within this repeated convolution and express them in a sum:
%\textcolor{red}{Bill can you have a look here}
\begin{multline}
    \mathcal{F}_B[\Gamma(B, \{x_i, \tau_i,\theta_i\})](\xi)=\\
    \frac{1}{2^N\Pr(\{X_i=x_i\})}\sum_{j\in J}\frac{\lambda_j}{2^{\sum_i|j_i|}e^{\sum_i|j_i|\tau_i/T}}\delta\left(\xi+\sum_{i}j_i\mu\tau_i\right),
\end{multline}
where $J$ is the set of all strings of $\{-1, 0,1\}$ of length $N$, matching the number of measurements included in the set $\{X_i=x_i,\tau_i,\theta_i\}$:
\begin{align}
    J&=\{-1,0,1\}^{\circledast N\iffalse{-1}\fi},\label{def: J set}
\end{align}    
and $\lambda_j$ is a complex number \iffalse{sum of complex phases}\fi that is non-trivial to calculate in general, but can be reduced to $1$ by appropriate change of basis (as shown in Appendix \ref{app: deriving KPE}). The result is an array of at most $3^{N\iffalse{-1}\fi}$ scaled $\delta$ distributions. In the simplified case discussed in Sec.~\ref{sec: simplified case}, the optimal measurement policy results in high degrees of overlapping peaks, resulting in $2^i$ peaks. The values of $H(X)$ and $H(X|B)$ are also readily computable from the Fourier domain, specifically:

\begin{align}
    H(X) &= -\sum_{X=x_i}\Pr(X=x_i)\ln(\Pr(X=x_i))\\
    \Pr(X=x_i) &= \mathcal{F}_B[\Pr(X=x_i|B)]\circledast\mathcal{F}_B[\Pr(B)](0)\label{eq: entropy from fourier}\\
    H(X|B)&=\mathcal{F}_B[H_\text{p}(X|B)]\circledast\mathcal{F}_B[\Pr(B)](0)\label{eq: cond ent from fourier}
\end{align}
where $H_\text{p}(X|B)$ is the pointwise conditional entropy:
\begin{align}
    H_\text{p}(X|B) &= -\sum_{X=x_i} \Pr(X=x_i|B)\ln(\Pr(X=x_i|B)).
\end{align}
The Fourier transform of the pointwise conditional entropy has been analysed in Appendix \ref{app: monotonicity of fourier coefficients}. 

Maximising mutual information places two competing pressures on a given likelihood function: $H(X)$ should be high (\emph{i.e.,} $X$ is an unbiased measurement) and $H(X|B)$ should be low (\emph{i.e.,} pointwise, $\Pr(X=x_i|B)$ is maximally biased).

\iffalse Calculating mutual information for the second measurement onward, we can see a distinct measurement-to-measurement interaction, where previous measurement values and outcomes ($x$, $\tau$ and $\theta$) directly impact future measurement performance. Further analysis of this is provided in Sec.~\ref{sec: simplified case} and in Appendix \ref{app: deriving KPE}\fi

\iffalse
\begin{enumerate}
    \item Ramsey likelihood functions <COVERED PREV>
    \item Control Variables
    \item One shot measurement mutual information using likelihoods and priors
    \item Implementation pseudo-code
    \item Inter-measurement interaction (frequency halving)
\end{enumerate}
\fi

\section{Simulated Behaviour}\label{sec: Simulated Behaviour}
In this work we compare multiple methods for parameter selection; random selection (selecting $\tau$ and $\theta$ randomly from $[0,\tau_\text{max})$ and $[0,2\pi)$ respectively), myopic variance or entropy reduction, and the parameter calculation as discussed in Sec.~\ref{sec: simplified case} (and consistent with one method proposed in \cite{kitaevQuantumMeasurementsAbelian1996}). The algorithm by which measurement parameter selection and simulation is conducted is described with the following pseudo-code:
\begin{enumerate}
    \item For each simulation instance, a true value, $B_\text{true}$, is selected from a Gaussian distribution
    \item A prior $\Pr(B)$ is defined
    \item For each simulated measurement $n$
    \begin{enumerate}
        \item Based on the prior, calculate the next measurement parameters $\tau_n, \theta_n$, either randomly from a uniform distribution (in the case of the Random method), by minimising the expected variance or entropy (in the case of the variance and myopic entropy methods) or in accordance with \iffalse Eqs.~\fi\eqref{def: adaptive tau} and \eqref{def: adaptive theta} (in the case of KPE method).
        \item Based on the true value of $B_\text{true}$ a measurement outcome is generated with \begin{align}
            \Pr(X=x_i) = \Pr(X=x_i|B=B_\text{true})
        \end{align}
        \item The posterior $\Pr(B|X=x_n,\tau_n,\theta_n)$ is calculated using Bayes' rule. Any relevant properties (posterior entropy and standard deviation) are calculated. $\Pr(B|X=x_n,\tau_n,\theta_n)$ is then used in place of $\Pr(B)$ and the process is repeated from step 2
    \end{enumerate}
\end{enumerate}

\begin{figure}[h]
    \centering
    \begin{subfigure}{0.36\textwidth}%48
            \centering
            \includegraphics[width=1.1\textwidth]{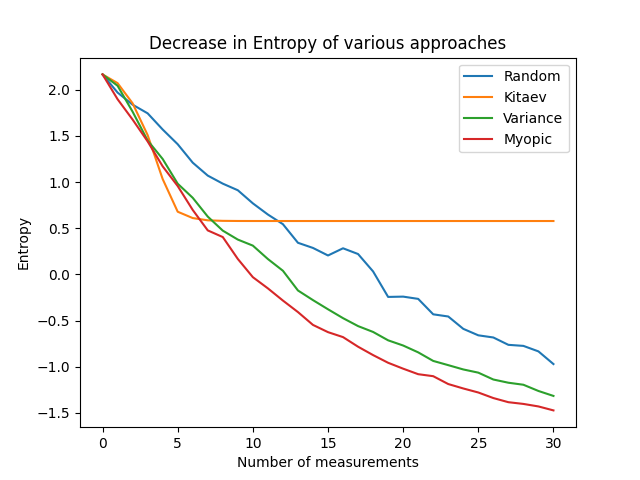}
    \end{subfigure}\quad
    \begin{subfigure}{0.36\textwidth}%48
            \centering
            \includegraphics[width=1.1\textwidth]{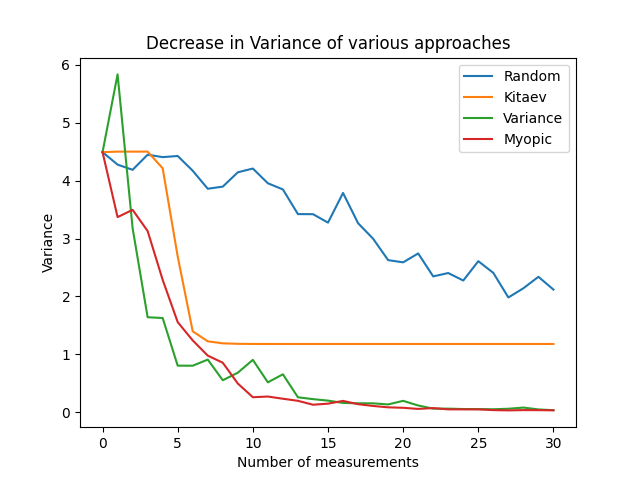}
    \end{subfigure}
    \caption{Average decrease in entropy (top) and variance (bottom) of a variable initialised with Gaussian prior (measured in nits) over 30 measurements with parameters selected using the KPE algorithm (orange), variance reduction (green), myopic mutual information maximisation (red) and randomly (blue). $T'$ is set to 10, $\tau'$ values are capped between $0$ and $5$. Averages are obtained over 8 separate realisations with differing $B$ values\iffalse{, standard deviations are marked every 5th step and staggered for clarity}\fi.}
    \label{fig: KPE vs myopic graphs}
\end{figure}

The results yielded by this simulation are presented in Fig.~\ref{fig: KPE vs myopic graphs}, showing the posterior variance and entropy reduction of difference measurement sequences.

\section{Analytic description of the simplified case}\label{sec: simplified case}

We can define a simplified case to allow for complete analytic description. Consider the case of measurements of a variable initialised with a diffuse prior ($k\rightarrow\infty$), in the absence of decoherence ($T\rightarrow\infty$). In this case any posterior that is ever obtained only contains periodic components introduced by prior measurements. This means \iffalse Eqs.~\fi\eqref{eq: entropy from fourier} and \eqref{eq: cond ent from fourier} can be evaluated directly:

\begin{align}
    \Pr(X=0) &= \mathcal{F}_B[\Pr(X=x_i|B)]\circledast\mathcal{F}_B[\Pr(B)](0)\\
    &=\frac{\mathcal{F}_B[\Pr(B)](0)}{2}+\frac{e^{-i\theta}\mathcal{F}_B[\Pr(B)](-2\mu\tau_{i+1})}{4}\nonumber\\&\quad+\frac{e^{i\theta}\mathcal{F}_B[\Pr(B)](2\mu\tau_{i+1})}{4}\\
    &=\frac{1}{2}+\frac{e^{-i\theta}}{4}\mathcal{F}[\Gamma(B, \{x_i, \tau_i,\theta_i\})\Pr(B)](2\mu\tau_{i+1})\nonumber\\
    &\quad+\frac{e^{i\theta}}{4}\mathcal{F}[\Gamma(B, \{x_i, \tau_i,\theta_i\})\Pr(B)](-2\mu\tau_{i+1})
\end{align}
The first term is constant, due to the normalisation of $\Pr(B)$, the latter two terms however are zero, unless $\Gamma$ contains a frequency component at $\xi =2\mu\tau_{i+1}$. This is equivalent to the statement: 
\begin{align}
    \exists j'_i\in J \text{ s.t. } \sum_ij'_i\tau_i=\tau_{i+1}\label{eq: mathing freq comps}
\end{align} (for $J$ as defined in \eqref{def: J set}). We can combine these contributing sequences into a new set $J'$, that contains only sequences $j_i$ such that the condition in \eqref{eq: mathing freq comps} is true, allowing the probability to be expressed:
\begin{align}
    \Pr(X=0)&=\frac{1}{2}+\frac{1}{2}\Re\left[e^{i\theta}\sum_{j'\in J'}\frac{\lambda_j'}{2^{\sum_i|j'_i|}e^{\sum_i|j'_i|\tau_i}}\right]
\end{align}
This means bias can be minimised in two ways: either $\tau_{i+1}$ can be selected such that $\lnot\exists j'\in J$ s.t. $\sum_ij'_i\tau_i=\tau_{i+1}$ (\emph{i.e.}, $J'=\varnothing$), or the second term must be entirely imaginary (\emph{i.e.}, $e^{i\theta}$ can be chosen such that $\theta = -\text{arg}\left[\sum_{j'\in J'}\frac{\lambda_j'}{2^{\sum_i|j'_i|}e^{\sum_i|j'_i|\tau_i}}\right]$). As shown in Appendix ~\ref{app: deriving KPE}, in the context that $\tau=q\tau_0$ for $q\in\mathbb{Q}$, a rezeroing of the $B$ variable ($B'=B+B_0$) can be chosen such that $\forall j\in J,\,\lambda'_j=1$. 

There are thus two regimes that maximise measurement entropy ($H(X)$): off-resonance measurements ($\lnot\exists j\in J\text{ s.t. }\tau_{i+1}=\sum_ij_i\tau_i$), and off-phase measurements ($\exists j\in J\text{ s.t. }\tau_{i+1}=\sum_ij_i\tau_i$, $\theta = -\text{arg}\left[\sum_{j'\in J'}\frac{\lambda_j'}{2^{\sum_i|j'_i|}e^{\sum_i|j'_i|\tau_i}}\right]$). In the off-resonance regime, we rely on the periodic components of the prior populating the high likelihood probability ($\Pr(X=x_i|B)>0.5$) and low likelihood probability ($\Pr(X=x_i|B)<0.5$) equally. In the off-phase regime, we locate the periodic components of the prior primarily within low likelihood bias regime ($\Pr(X=x_i|B)\approx0.5$). notably in the presence of decoherence, there is another high entropy regime, $\tau\rightarrow\infty$ where the entire likelihood probability is a low bias.

\begin{figure}
    \centering
    \includegraphics[width=0.5\linewidth]{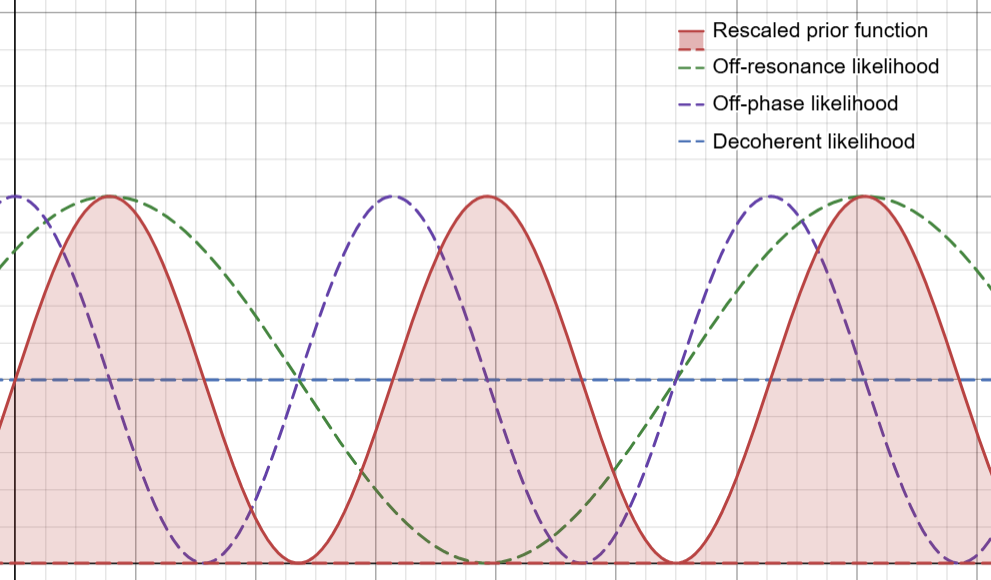}
    \caption{The alignment between a rescaled prior (red) with periodic components, compared against off-resonant likelihood (green), off-phase likelihood (purple) and fully decohered likelihood (blue). All three of these measurement likelihoods yield a low bias and a high measurement entropy.}
    \label{fig:non-biased likelihoods}
\end{figure}

To maximise mutual information, we also need to minimise conditional entropy ($H(X|B)$). Using \iffalse \iffalse Eq.~\fi\fi\eqref{eq: cond ent from fourier} in the simplified context ($T\rightarrow\infty$, $k\rightarrow\infty$):
\begin{align}
    H(X|B)&=\mathcal{F}_B[H_\text{p}(X|B)]\circledast\mathcal{F}_B[\Gamma(B, \{x_i, \tau_i,\theta_i\})\Pr(B)](0)\\
    &=\sum_{j\in J}\sum_{i'\in I'_j} \alpha_{i'} e^{i\theta}\frac{\lambda_j}{2^{\sum_i|j_i|}e^{\sum_i|j_i|\tau_i/T}},\label{eq: conditional entropy fourier series}
\end{align}
where $I'_j$ is the set of integers $i' = 2\sum_ij_i\frac{\tau_i}{\tau_{i+1}}$, and $\alpha_{i'}$ is a discrete series that is negative and monotonically increasing for $i'>0$ and positive for $i'=0$ (as shown in Appendix \ref{app: monotonicity of fourier coefficients}). We can break \iffalse \iffalse Eq.~\fi\fi\eqref{eq: conditional entropy fourier series} into the zero frequency term and the non-zero terms, noting that the zero frequency term is independent of measurement parameters ($\tau_{i+1}$, $\theta_{i+1}$), yielding:
\begin{align}
    H(X|B)&= \alpha_0 + \sum_{j\in J^*}\sum_{i'\in I^*_j} \alpha^*_{i'} e^{i\theta}\frac{\lambda_j}{2^{\sum_i|j_i|}e^{\sum_i|j_i|\tau_i/T}},%\label{eq: conditional entropy fourier series}
\end{align}
where $J^*, I^*, \alpha^*$ are the equivalent sets to $J$, $I'$, $\alpha$ with their zero elements removed. The zero/non-zero split allows for simplification. $\alpha_0$ is independent of $\tau_{i+1}$, $\theta_{i+1}$, the double sum is the element-wise dot product of a monotonic series ($\alpha^*_{i'}$) and the Fourier coefficients of the prior with which they match.

The optimal behaviour can be calculated by induction on the initial prior. In the $i=0$ case, $\Gamma(B)=1$, thus \iffalse \iffalse Eq.~\fi\fi\eqref{eq: conditional entropy fourier series} reduces to $H(X|B)=\alpha_0$. In the second case ($i=1$) however $\Gamma(B)=\Pr(X=x_1|B,\tau_1, \theta_1)$, thus there is a single periodic component to contribute to the sum in \iffalse \iffalse Eq.~\fi\fi\eqref{eq: conditional entropy fourier series}, arising from the $J=1,\,-1$ peaks:

\begin{align}
    H(X|B)&= \alpha_0 + \sum_{j\in \{1,\,-1\}}\sum_{i'\in I^*_j} \alpha^*_{i'} e^{i\theta_2}\frac{e^{i(\theta_1+\frac{\pi}{2} x_1)}}{2e^{\tau_1/T}}\\
    &=\left\{ 
    \begin{array}{cc}
        \alpha_0 + \cos(\theta_2-(\theta_1+\frac{\pi}{2} x_1))\alpha_1 & \tau_2 = \frac{1}{2}\tau_1\\
        \alpha_0 & \tau_2\neq\frac{1}{2}\tau_1
    \end{array}\right.
\end{align}
Conditional entropy is then minimised when 
\begin{gather}
    \tau_2 = \frac{1}{2}\tau_1\label{def: tau iteration i=2}\\
    \cos(\theta_2 - \theta_1 - \frac{\pi}{2}x_1) = 1\iff \theta_2 = \theta_1+\frac{\pi}{2}x_1.\label{def: theta iteration i=2}
\end{gather} 
Notably, these restrictions on $\tau$ and $\theta$ are consistent with the maximisation of the measurement entropy $H(X)$, meaning that satisfying \iffalse Eqs.~\fi\eqref{def: tau iteration i=2} and \eqref{def: theta iteration i=2} is sufficient to maximise mutual information for the second measurement $I(X_2;B|X_1=x_1)$.

As shown in Appendix \ref{app: deriving KPE} we can continue this argument inductively for $i>2$, leveraging the monotonicity of the Fourier coefficients, both of the sequence $\alpha^*$ and the $\Gamma$ coefficient when the myopic maximisation process is followed. This leads to the complete adaptive parameters in the iterative form:

\begin{align}
    \tau_i &= \frac{1}{2}\tau_{i-1}\label{def: adaptive tau}\\
    \theta_i &= \frac{\theta_{i-1}+\pi x_{i-1}}{2}.\label{def: adaptive theta}
\end{align}
This amounts to an exponential decrease in exposure time $\tau$ for each subsequent Ramsey measurement, with a readout phase $\theta$ that depends on the prior phase and outcome. If the interaction with the magnetic field for some final time $\tau_\text{final}$ is thought of as a physical operator, the exponential reduction of exposure time $\tau$ with and adaptive $\theta$ responding to previous results ($x_{i-1}$) is analogous to prior results. In a sensing protocol discussed in Ref.~\cite{kitaevQuantumMeasurementsAbelian1996} and posed explicitly (in an interferometric context) in Ref.\cite{higginsDemonstratingHeisenberglimitedUnambiguous2009}, individual qubits are exposed to a unitary operator an exponentially reducing amount of times, with an adaptive unitary rotation applied pre-measurement.

\iffalse
\begin{enumerate}
    \item Analytic description of entropy minimisation in the absence of decoherence and prior minimisation 
    \item Analytic minimisation conditional entropy $S(X_i|a,\theta, \{X\})$ of frequency halving
    \item Analytic Maximisation of measurement entropy $S(X|\{X\})$ (no bias theorem)
\end{enumerate}
\fi

\section{Conclusion}

In this work we have presented a myopic entropy based approach for NV magnetometry, demonstrated how it can be calculated in general, and described its exact behavior in a simplified case finding it consistent with pre-existing algorithms used in different contexts \cite{kitaevQuantumMeasurementsAbelian1996, higginsEntanglementfreeHeisenberglimitedPhase2007, higginsDemonstratingHeisenberglimitedUnambiguous2009, degenQuantumSensing2017, barrySensitivityOptimizationNVdiamond2020}. We have simulated the performance of the myopic entropy reduction algorithm outside of the regime described in Sec.~\ref{sec: simplified case}, finding the myopic approach provided advantage over the simplified approach in the long sequence limit.

Entropy minimisation offers a straightforward conceptual framework for adaptive magnetometry protocols. Due to the submodularity of the entropy minimisation problem,  myopic approximations reduce entropy reduction by at most a constant scalar factor ($1-1/e\approx 0.63$). As shown in Sec.~\ref{sec: Simulated Behaviour}, in the presence of decoherence myopic entropy reduction on a localised prior is capable of functioning in the large measurement limit beyond the point the analytic relationship shown in \iffalse Eqs.~\fi\eqref{def: adaptive tau} and \eqref{def: adaptive theta} breaks down. In the context of real time adaptive sensing, direct computation and minimisation of posterior entropy is slower than of a Ramsey measurement, which happens on the order of tens of microseconds \cite{bonatoOptimizedQuantumSensing2016}. In practical sensing applications this leads to some degree of downtime between measurements, and lowering both the overall measurement frequency and sensitivity as a function of exposure time. This problem is by no means insurmountable and analytic methods exist that offer simplifications to the process. The sinusoidal structure of the likelihood functions shown in \iffalse Eqs.~\fi\eqref{def: pr(x=0) likelihood function} and \eqref{def: pr(x=1) likelihood function}, and the spin precession process more generally \cite{degenQuantumSensing2017, barrySensitivityOptimizationNVdiamond2020}, allows for efficient analysis in terms of Fourier coefficients as discussed in Ref.~\cite{berryHowPerformMost2009} and Appendices \ref{app: deriving KPE}-\ref{app: monotonicity of fourier coefficients}. It is not guaranteed that the measurement parameters can be calculated as simply as \iffalse Eqs.~\fi\eqref{def: adaptive tau} and \eqref{def: adaptive theta}, however analytic calculation of $\tau$ and $\theta$ are likely possible without explicit calculation of the posteriors or searching over the entire $\tau, \theta$ space. 

In this work we have modeled an idealised system, in which the only source of noise arises from decoherence. Many sources of error have been well discussed in the literature \cite{degenQuantumSensing2017, barrySensitivityOptimizationNVdiamond2020}, notably errors arising from laser decoherence, detuning, pulse length error, and readout (\emph{e.g.}, photon detection error). Each of these error sources can be modeled by changes to the likelihood function, either in a manner consistent with the discussion in Appendix \ref{app: phase descriptions} or requiring a more holistic physical derivation. The myopic entropy reduction method is still applicable in these cases, although analytic derivations as shown in Appendices \ref{app: phase descriptions}-\ref{app: monotonicity of fourier coefficients} would need to be expanded upon in future investigation. Dephasing error can be included into the analysis covered in Appendix \ref{app: monotonicity of fourier coefficients}, more generally, changes to the prior and likelihood functions would require new derivations following similar structure to Appendix \ref{app: deriving KPE}.

The posterior entropy reduction model posits to simultaneously maximises the informative value of the measurement through maximisation of $H(X)$, while minimising the amount of that informative value that is unrelated to the estimation problem at hand through the minimisation of $H(X|B)$. Lowering conditional entropy demands that the probability is maximally biased over regions where $\Pr(B)$ is large. Maximising measurement entropy demands minimising the overall bias of the measurement, whilst this can be achieved by reducing the bias of $\Pr(X=x_i|B)$ pointwise, this conflicts with the requirement imposed by minimisation of $H(X|B)$. The increase in $H(X)$ from reduction in pointwise bias of the likelihood is measurement of a random variable unrelated to the reduction of the prior $\Pr(B)$, it is met with an equal increase in $H(X|B)$ and thus has no impact on $I(X;B)$. The alternative method to increase $H(X)$ is to balance regions of high and low bias in equal measure over the prior. This requires either a high slope of the likelihood function (consistent with the linear scaling regime described in Ref.~\cite{degenQuantumSensing2017}) or separation of the prior probability into separated regions (as is present in the low decoherence case).

The conceptual simplicity of the entropy reduction algorithm allows it to be applied even for more complex forms of the likelihood function. Likelihood functions more uniquely tuned to a given experimental set up offer tuned measurement protocols optimised for the specific parameters and error models most applicable to the situation. This again stresses the value of analytic simplification of a more general form than provided in Appendix \ref{app: deriving KPE}.

\appendices
\section{A physical description of the Ramsey experiment}\label{app: phase descriptions}
\iffalse
\begin{itemize}
    \item Physical set up - B field and Laser Hamiltonian
    \item Setting up the Bloch Sphere
    \item The rotating wave 
    \item What is a $\pi/2$ pulse
    \item Controlling relative phase ($\theta$)
    \item Additional considerations (off resonance absorption, $\pi/2$ pulse inaccuracy, decoherence in the laser, photon collection)
\end{itemize}
\fi
There are multiple descriptions of the mechanics behind the Ramsey measurement \cite{degenQuantumSensing2017, barrySensitivityOptimizationNVdiamond2020, craigieResourceefficientAdaptiveBayesian2021, foot_atomic_2004, akkermans_atom-photon_nodate, shiga_locking_2012}. We present a short treatment here to ground the concepts we rely on to reach \iffalse Eqs.~\fi\eqref{def: pr(x=0) likelihood function} and \eqref{def: pr(x=1) likelihood function} and to collate and discuss additional effects not included in any those analyses individually.

\subsection{The NV Hamiltonian}\label{app: NV ham}

Here we present a treatement of the NV centre, following closely the work done in Ref.~\cite{barrySensitivityOptimizationNVdiamond2020} Sec.~1c and Ref.~\cite{akkermans_atom-photon_nodate} Ch.~2, covering the NV Hamiltonian and optically pumped two level atoms respectively. We also extend this treatment slightly to explicitly cover the measurement phase term $\theta$ as seen in \iffalse Eqs.~\fi\eqref{def: pr(x=0) likelihood function} and \eqref{def: pr(x=1) likelihood function}, and how phase in th elikelihood function arises from the physical phase of the optical field \cite{akkermans_atom-photon_nodate, shiga_locking_2012}.

The NV centre Hamiltonian, under external magnetic fields, can be broken down into three components, the magnetic interaction term, nuclear spin term, and electric strain term \cite{barrySensitivityOptimizationNVdiamond2020}:
\begin{align}
    H &= H_\text{Mag} +H_\text{Nuc}+H_\text{ES}.
\end{align}

Each of these components can be written individually, starting with the magnetic interaction term that governs the interation between the bound electron, external magnetic field and zero field splitting:
\begin{align}
    H_\text{Mag} &= hDS_z^2+g_e\mu_B(\vec{B}\cdot\vec{S})
\end{align}
where $D$ is the zero field splitting parameter $\approx$ 2.87 GHz, $\vec{S}$ is the spin vector of the bound electron (with $S_z$ aligned along the N-V axis). $g_e\approx 2.003$ is the NV electronic g-factor, $\mu_B$ is the Bohr magneton and $h$ is Planck's constant, these can be combined to a singular $\mu$ term, as shown in \iffalse \iffalse Eq.~\fi\fi\eqref{def: nv hamiltonian}. The nuclear term governs interactions between the Nitrogen nuclear  and electron spins and the nitrogen and magnetic field:
\begin{multline}
    H_\text{nuc} = hA_{||}S_zI_z + hA_{\perp}(S_xI_x+S_yI_y) + hP(I^2_z-I(I+1)/3)\cr-g_N\mu_N(\vec{B}\cdot\vec{I}),
\end{multline}
where $A_{||}$ and $A_{\perp}$ are the axial and transverse hyperfine coupling constants (on the order of 2-3 MHz depending on the nitrogen isotope), $P$ is the electric quadrapole moment ($\approx2$MHz for NV$^{14}$ and ignorable for NV$^{15}$) \cite{barrySensitivityOptimizationNVdiamond2020}, $\vec{I}$ is the nuclear spin vector ($I_{\text{NV}^{14}}$=1/2 and $I_{\text{NV}^{15}}$=1), and $g_N$ and $\mu_N$ are the nuclear electronic g-gactor and nuclear magneton respectively.
Finally, $H_\text{ES}$ captures the stress and strain of the electron within the diamond lattice. These terms are complex, but small in comparison to the other two terms in the Hamiltonian, and thus can be ignored for this treatment, though a more complete description can be found in Ref.~\cite{barrySensitivityOptimizationNVdiamond2020}.

\subsection{The interaction Hamiltonian}\label{subapp: interaction ham}

When exposed to an oscillating magnetic field (\emph{e.g.,} a laser pulse) it is convenient to separate the magnetic field into a static field associated with the DC field and the oscillating field, we can thus write:
\begin{align}
    H_\text{mag}=hDS_z^2+g_e\mu_bB_zS_z+H_\text{int}
\end{align}
where $H_\text{int}$ is the interaction with the optical field. If we treat the electric field classically, we can write the interaction term:
\begin{align}
    H_\text{int}&=-\vec{d}\cdot\vec{E}
\end{align}
where $\vec{d}$ is the dipole unit vector and $E$ is the electric field \cite{akkermans_atom-photon_nodate}. We can express $\vec{d}$ in terms of the ground and excited states:
\begin{align}
    \vec{d} &= \left(\ket{0}\bra{0}+\ket{1}\bra{1}\right)\vec{d}\left(\ket{0}\bra{0}+\ket{1}\bra{1}\right)\\
    &=\bra{0}\vec{d}\ket{1}\ket{0}\bra{1}+\bra{1}\vec{d}\ket{0}\ket{1}\bra{0}
\end{align}
Where here we've defined $\ket{0}$ as the ground state and $\ket{1}$ as the excited state (up to a phase). It is often useful to select a phase for the excited state such that $\bra{0}\vec{d}\ket{1}=\bra{1}\vec{d}\ket{0}\in\mathbb{R}$ \cite{akkermans_atom-photon_nodate}. In this case, we can write the interaction Hamiltonian:
\begin{align}
    H_\text{int} &=-\left( \bra{0}\vec{d}\ket{1}\ket{0}\bra{1}+\bra{1}\vec{d}\ket{0}\ket{1}\bra{0}\right)\nonumber\\&\quad\cdot\left(\frac{E_0e_z}{2}\cos(\omega t)\sigma_z\right)\\
    &=-\bra{0}\vec{d}\cdot\hat{\vec{e}}\ket{1}\frac{E_0}{2}\cos(\omega t)(\sigma_x\cdot\sigma_x)\\
    &=h\Omega_R \cos(\omega t)\sigma_y,\label{eq: real value H int}
\end{align}
where $\hat{\vec{e}}$ and $e_z$are the component of the electric field unit vector and the component aligned with the dipole respectively, $E_0$ is the magnitude of the external oscillating electric field, and $\Omega_R=\frac{1}{h}\bra{0}\vec{d}\cdot\hat{\vec{e}}\ket{1}\frac{E_0}{2}$ is the Rabi frequency \cite{akkermans_atom-photon_nodate}. This is the Hamiltonian that generates excitation of the ground state into superposition. We have however implicitly linked the definition of the excited state to the phase of the laser pulse, both in the selection of the gauge of the electromagnetic field and in the definition of the dipole moment \cite{akkermans_atom-photon_nodate}. This means that, if exposed to a laser with differing phase a more complex definition of $H_\text{int}$ is required. Specifically, consider $\ket{1'}=e^{-i\theta}\ket{1}$ such that:

\begin{align}
    H_\text{int} &= -\left( \bra{0}\vec{d}'\ket{1'}\ket{0}\bra{1'}+\bra{1'}\vec{d}'\ket{0}\ket{1'}\bra{0}\right)\nonumber\\&\quad\cdot\left(\frac{E_0e_z}{2}\cos(\omega t)\sigma_{z'}\right)\\
    &=-\left(\bra{0}\vec{d}\ket{1}e^{i\theta}\ket{0}\bra{1}+\bra{1}\vec{d}\ket{0}e^{-i\theta}\ket{1}\bra{0}\right)\nonumber\\&\quad\cdot\left(\frac{E_0e_z}{2}\cos(\omega t)\sigma_{z}\right)
    \\
    &=-\bra{0}\vec{d}\ket{1}\left(e^{i\theta}\cos(\theta)\sigma_x-\sin(\theta)\sigma_y\right)\nonumber\\
    &\quad\cdot\left(\frac{E_0e_z}{2}\cos(\omega t)\sigma_{z}\right)\\
    &=h\Omega_R\cos(\omega t)(\cos(\theta)\sigma_y+\sin(\theta)\sigma_x)
\end{align}
We thus have a tunable axis of rotation that can be targeted by an external laser pulse. Importantly this rotation is in Bloch space, and not a physical rotation in real space. One way to implement this experimentally is to expose the NV centre to a coherent laser pulse initially and re-expose the system to the same laser pulse (maintaining coherence) with addition of a phase defect or change in path length \cite{shiga_locking_2012}. In doing so, it is crucial that the coherence of the laser is maintained not just over the pulse length, but also the exposure length of the NV centre to the external field. Lack of laser coherence will lead to an uncertainty in the effective value of $\theta$.

Overall, we can approximate the Hamiltonian  of the system only in terms of the static magnetic field and atom-field interaction, emitting the zero field splitting, nuclear terms, and electron stress terms are multiple orders of magnitude smaller, thus:
\begin{align}
    H \approx \begin{cases}
        \frac{\hbar\mu}2 B \sigma_z & E_0=0\\ 
        \frac{\hbar\mu}2 B \sigma_z + h\Omega_R\cos(\omega t)(\cos(\theta)\sigma_y+\sin(\theta)\sigma_x) &E_0\neq 0
    \end{cases}.
\end{align}
This removal of small energy differences between energy levels leads to additional decoherence within the 2 level subsystem, which broadens laser absorption peaks and limits the fidelity of the sensor\cite{degenQuantumSensing2017, barrySensitivityOptimizationNVdiamond2020}.

\subsection{Bloch sphere visualisation and geometric interpretation of errors}

The Bloch sphere provides an effective way to visualise
the evolution of the state. Since the NV$^-$ centre is not truly a two level system, not all phenomena can be captured (\emph{e.g.}, interactions with the nuclear degrees of freedom). 

The $z$ axis of the Bloch sphere corresponds to the weighting of the state between the ground state $\ket{0}$ and the excited state $\ket{1}$. There is an additional rotational degree of freedom (where the $\ket{+}$ state is defined around the circumference), this amounts to the choice of phase discussed in Sec.~\ref{subapp: interaction ham}, and the we can define $\ket{+}$ to be the state reached when a given sample laser pulse is applied to the ground state with magnitude to enact a $\pi/2$ pulse (this is analogous to the choice of $\ket{1}$ that leads to \iffalse Eq.~\fi\eqref{eq: real value H int}). This definition of $\ket{+}$ also implicitly defines a reference phase for pumping lasers.

The $\pi/2$ pulse implements a global rotation unitary, notably this maps two points on the circumference of the Bloch sphere to the poles on the $z$ axis. These two points are then perfectly distinguishable, and a choice of $\theta$ allows for selection of an axis of optimal sensitivity. Any given rotation also has fixed points, these lie on the circumference and thus correspond to points of zero bias (mapping to equal probability of emission and non-emission). To generate the probabilities shown in \iffalse Eqs.~\fi\eqref{def: pr(x=0) likelihood function} and \eqref{def: pr(x=1) likelihood function}, we consider the Ramsey measurement to include 1) an initial $\pi/2$ pulse from the ground state, 2) a free procession around the $z$ axis caused by the magnetic field, and 3) a $-\pi/2$ pulse that rotates out of the circumference, in lead up to a projective measurement in the $z$ direction.

There are a two sources of error that can be visual well within the Bloch sphere framework, $\pi/2$ pulse strength and $\theta$ error. Since the magnitude of the rotation induced by a pulse depends on the magnetic field strength, which is unknown (for $B_T=B_0+B$), the true exposure time and amplitude needed to implement a $\pi/2$ rotation are not known exactly. Practically speaking, it is typically sufficient to treat the total magnetic field as the same as the bias field, $B_T\approx B_0$. This can lead to an under or overshoot from the circumference. Because the sensitivity depends on the transverse distance from the $z$ axis, errors are geometrically reduced by the lack of curvature around the circumference. $\theta$ error, as may arise from noise in the laser phase changes the axis of rotation. This creates a set of different rotations for any given point. This error is capable of completely shifting the outcome probability, however, for small values of $\Delta\theta$ ($\Delta\theta<\pi/2$) error only functions to reduce to overall contrast.

\begin{figure}
    \centering
    \begin{subfigure}{0.24\textwidth}%31
            \centering
            \includegraphics[width=1\textwidth]{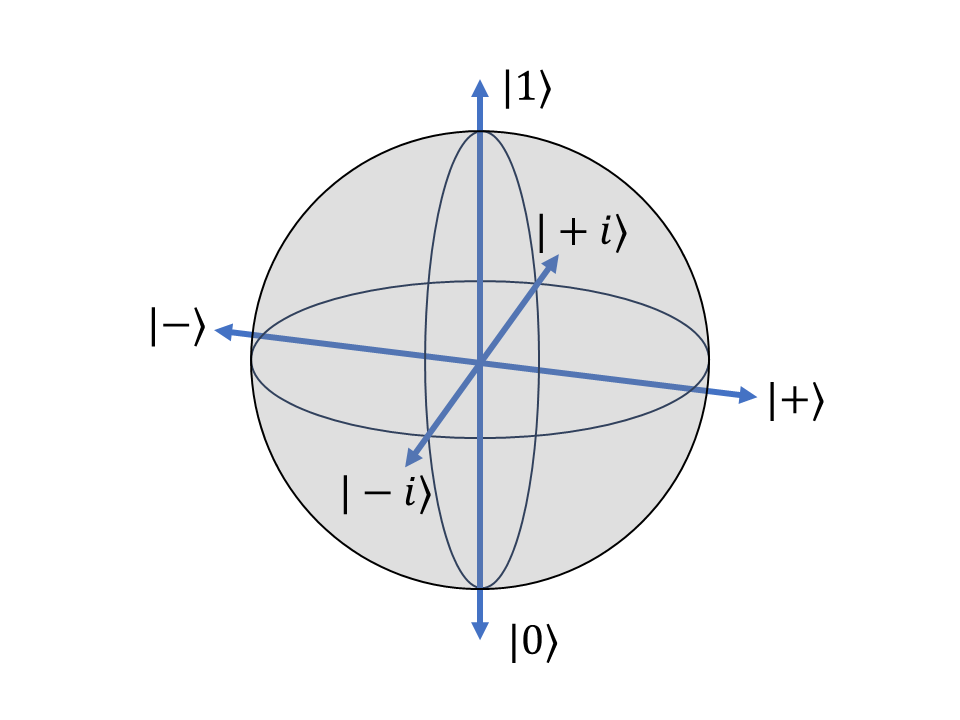}
    \end{subfigure}
    \begin{subfigure}{0.24\textwidth}%31
            \centering
            \includegraphics[width=1\textwidth]{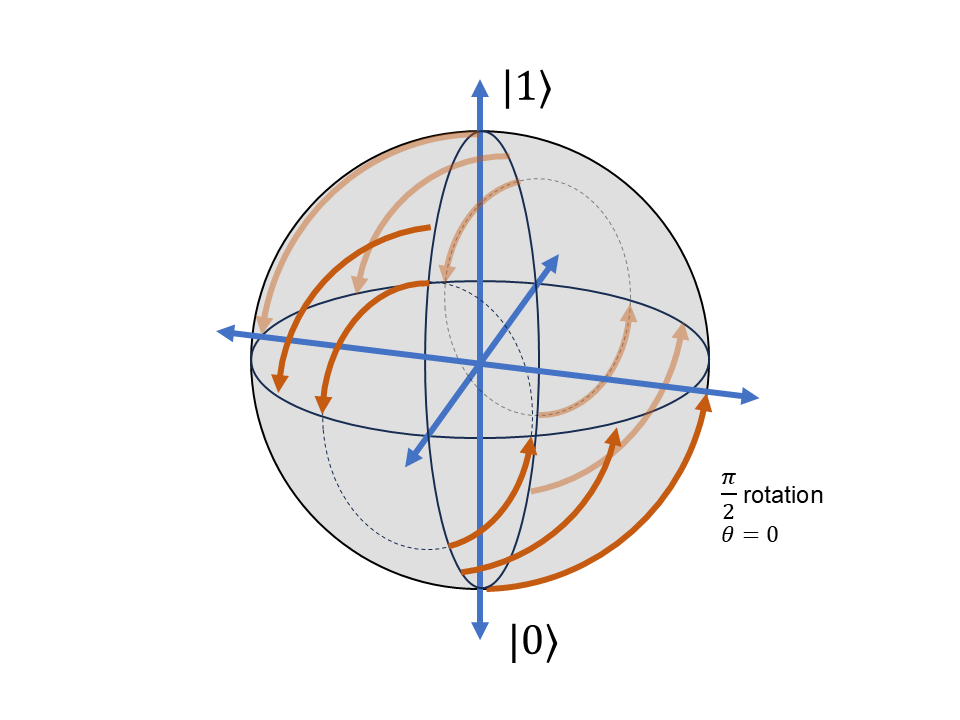}
    \end{subfigure}
    \begin{subfigure}{0.24\textwidth}%31
            \centering
            \includegraphics[width=1\textwidth]{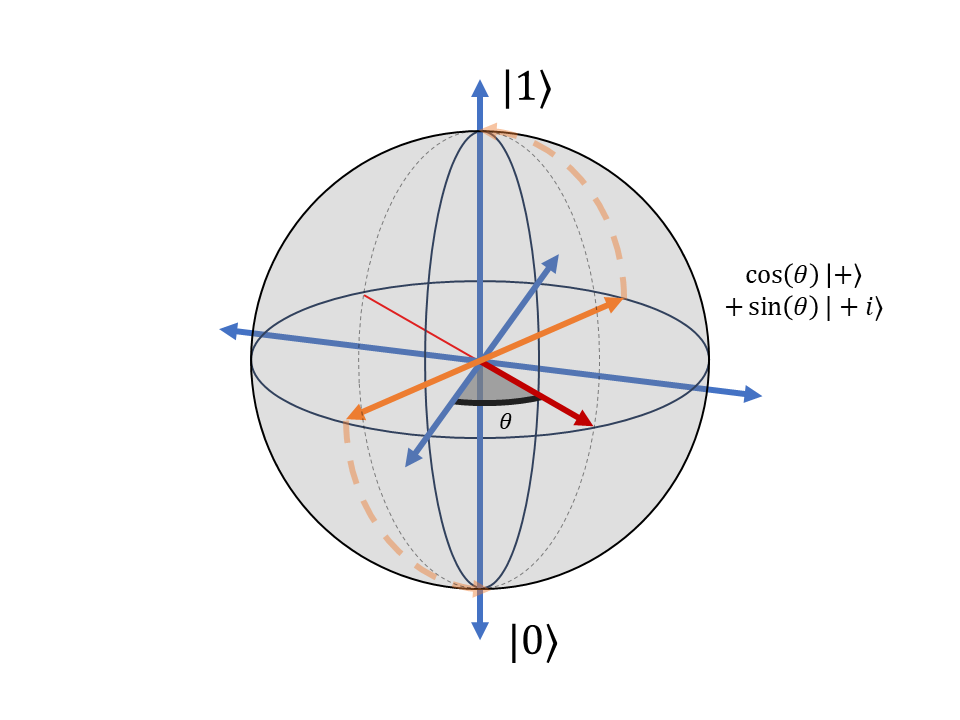}
    \end{subfigure}
    
    \begin{subfigure}{0.24\textwidth}%31
            \centering
            \includegraphics[width=1\textwidth]{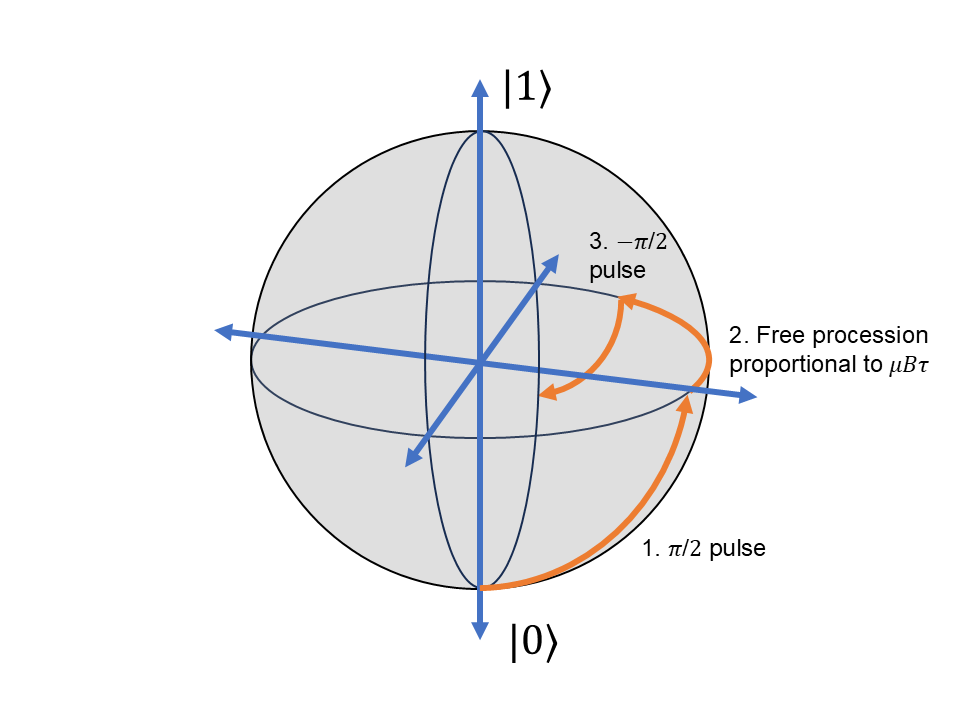}
    \end{subfigure}
    \begin{subfigure}{0.24\textwidth}%31
            \centering
            \includegraphics[width=1\textwidth]{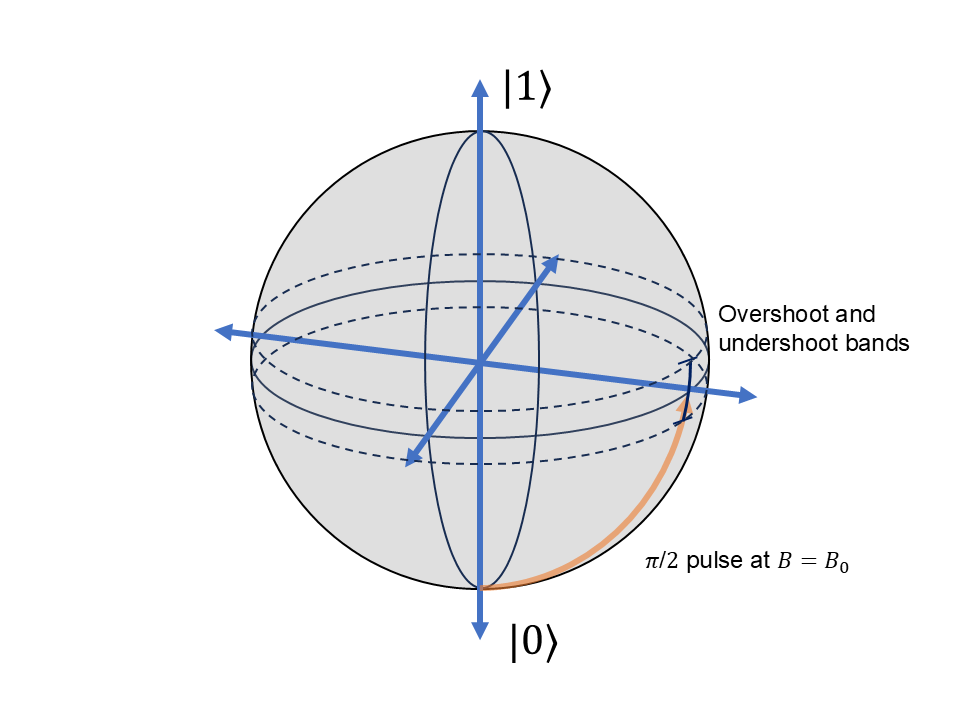}
    \end{subfigure}
    \begin{subfigure}{0.24\textwidth}%31
            \centering
            \includegraphics[width=1\textwidth]{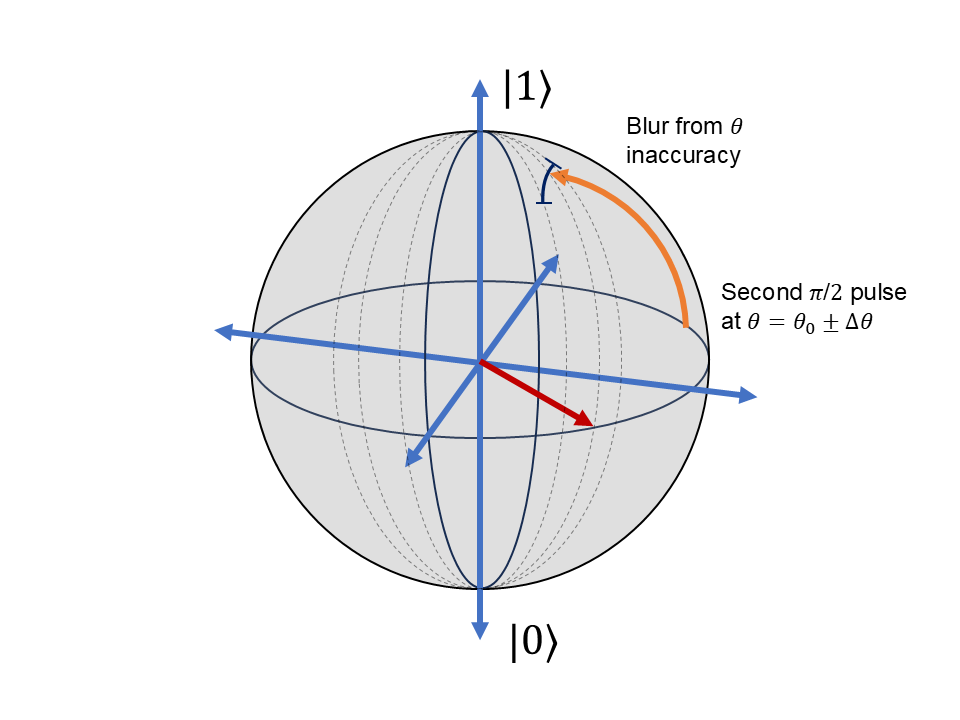}
    \end{subfigure}
    \caption{Bloch sphere representation for the two level model of the NV centre. Top to bottom: 1. the Bloch sphere, with unit vectors marked; 2. the rotation generated by a $\pi/2$ laser pulse (with $\theta=0$); 3. the Axis of optimal sensitivity (orange) for a given readout phase offset $\theta$, with generating rotational axis marked (red); 4. the three actions that make up a single Ramsey measurement sequence; 5. the error bands that arise from larger or smaller strength of the $\pi/2$ pulse; and 6. the error arising from blurring in $\theta$ relative phase.}
    \label{fig:NV bloch diagrarms}
\end{figure}

\iffalse\subsection{Projection and likelihood functions}

Since we assume readout happens via a projection along the $z$ axis (\emph{i.e.}, arising from measurements of the excited state of the electron in the field), the probabilities of detecting a ground or exited state can be written:
\begin{align}
    \Pr(X=0|\psi) &= \left|\braket{0\mid\psi}\right|^2\\
    &=\frac{1-s_z}{2}\\
    \Pr(X=1|\psi) &= \left|\braket{1\mid\psi}\right|^2\\
    &=\frac{1+s_z}{2},
\end{align}
where $s_z$ is the component of the state Bloch vector (after the second readout pulse). We can track \vec{s} throughout the evolution explicitly:
\begin{align}
    \vec{s} &= (0,0,-1)\\
    \intertext{After a $\pi/2$ pulse at reference phase:}
    \vec{s} &= (1,0,0)\\
    \intertext{After procession for time $\tau$ under field $B$ (roting that the rotating wave approximation allows for $B_0$ to be ignored:}
    \vec{s} &= (\cos(\mu\tau B),\sin(\mu\tau B), 0)\\
    \intertext{After a second readout pulse with relative phase $\theta$:}
    \vec{s} &= (
\end{align}
\fi

\section{Derivation of the mutual information from the likelihood}\label{app: Mutual information derivation}

To derive \iffalse Eq.~\fi\eqref{eq: mutual information from the prior}, we start by collecting basic Baysian definitions and basic Shannon entropic definitions \cite{nielsenQuantumComputationQuantum2010, coverThomasElementsofInformationTheory}. For some event $X$ with outcomes $\{x_i\}$, a continuous variable $B$, with a set of likelihood functions for each outcome $\Pr(X=x_i|B)$ and prior distribution for $B$, $\Pr(B)$. from there we can define the expected probabilities $\Pr(X=x_i)$, mutual information $I(X;B)$, entropy of the measurement $H(X)$ and conditional information $H(X|B)$:

\begin{align}
    \Pr(X=x_i) &= \int \Pr(X=x_i|B)\Pr(B)dB\label{def: bayesian probability}\\
    I(X;B) &= H(X) - H(X|B)\label{def: mutual information}\\
    H(X) &=-\sum_{x_i} \Pr(X=x_i)\ln\left(\Pr(X=x_i)\right)\label{def: shannon entropy}\\
    H(X|B) &= -\int \sum_{x_i}\Pr(X=x_i|B)\ln\left(\Pr(X=x_i|B)\right)\Pr(B)dB\label{def: conditional entropy}
\end{align}
We substitute \iffalse Eq.~\fi\eqref{def: bayesian probability} into \iffalse Eq.~\fi\eqref{def: shannon entropy} to obtain

\begin{align}
    H(X) &=-\sum_{x_i} \left(\int \Pr(X=x_i|B)\Pr(B)dB\right)\nonumber\\&\quad\ln\left(\int \Pr(X=x_i|B)\Pr(B)dB\right)
\end{align}
We then rewrite \iffalse Eq.~\fi\eqref{def: mutual information} solely in terms of the likelihoods and prior distribution:
\begin{multline}
    I(X;B) = -\sum_{x_i} \left(\int \Pr(X=x_i|B)\Pr(B)dB\right)\\\ln\left(\int \Pr(X=x_i|B)\Pr(B)dB\right) \\+ \int \sum_{x_i}\Pr(X=x_i|B)\ln\left(\Pr(X=x_i|B)\right)\Pr(B)dB.\label{eq: mutualinformation from likelihoods (appendix)}
\end{multline}

\section{Inductive calculation of optimal behaviour in the absence of decoherence and prior}\label{app: deriving KPE}

We show that optimal choice of measurement parameters follows the iterative form
\begin{align}
    \tau_i &= \frac{1}{2}\tau_{i-1}\\
    \theta_i &= \frac{\theta_{i-1}+\pi x_{i-1}}{2}
\end{align}
in the absence of decoherence ($T\rightarrow\infty$) over a diffuse prior ($k\rightarrow\infty$). We can show this inductively, by maximising the mutual information and equivalently by simultaneously maximising $H(X)$ and minimising $H(X|B)$. First we show that, if the previous measurement parameters are consistent with those generated by the KPE algorithm, the next measurement parameters generated by the myopic algorithm will agree with the KPE algorithm. We can also show trivially, under the assumption of diffuse priors, that any first measurement defines a specific KPE sequence.

Based on Ref.~\cite{higginsEntanglementfreeHeisenberglimitedPhase2007}, the KPE can be categorised as a measurement schedule in the following way:

\begin{align}
    \tau_n &= \frac{1}{2}\tau_{n-1}\label{def: KPE parameters iterative tau}\\
    \theta_n &= \frac{\theta_{i-1} + \pi x_{i-1}}{2}.\label{def: KPE parameters iterative theta}
\end{align}
This can be alternatively expressed as:

\begin{align}
    \tau_n &= \frac{1}{2^n}\tau_{0},\label{def: KPE parameters sequential tau}\\
    \theta_n &= \frac{1}{2^n}\theta_0 + \sum_{i=1}^{n-1}\frac{\pi x_{i}}{2^i},\label{def: KPE parameters sequential theta}
\end{align}
where $\tau_0$ and $\theta_0$ are two hyperparameters chosen prior to the experiment. We can see here that whilst $\theta_n$ depends non-trivially on the entire ordered list of previous measurement outcomes $\{x_i\}$, $\tau_n$ depends only on its position in the sequence and the initial parameter $\tau_0$ and thus is not adaptive in its own right.

\subsection{Fourier representation of features}

As shown in Sec.~\ref{sec: bayesian analysis}, the mutual information, likelihood, prior and posterior can all be represented in terms of the Fourier coefficients. Specifically:
\begin{align}
    \mathcal{F}_B\left[\Pr(B)\right](\xi) &= e^{-(\pi k 
    \xi)^2}\\
    \mathcal{F}_B\left[\Pr(X=x_i|B, \theta,\tau)\right](\xi) &= \frac{1}{4}\big(2\delta(\xi) \nonumber\\
    &\quad+ e^{-\tau/T-(2\pi  \mu\tau\theta +\pi x_i) i}\delta(\xi+2\mu\tau)\nonumber\\
    &\quad + e^{-\tau/T+(2\pi  \mu\tau\theta +\pi x_i)i}\delta(\xi-2\mu\tau)\big)\\
    \mathcal{F}_B\left[\Pr(B|X, \theta,\tau)\right](\xi)&= \frac{\mathcal{F}_B\left[\Pr(X|B, \theta,\tau)\right]\circledast\mathcal{F}_B\left[\Pr(B)\right](\xi)}{\mathcal{F}_B\left[\Pr(X|B, \theta,\tau)\right]\circledast\mathcal{F}_B\left[\Pr(B)\right](0)}\\
    H(X) &= -\sum_{X=x_i}\Pr(X=x_i)\ln(\Pr(X=x_i))\\
    \Pr(X=x_i) &= \mathcal{F}_B[\Pr(X=x_i|B)]\circledast\mathcal{F}_B[\Pr(B)](0)\\
    H(X|B)&=\mathcal{F}_B[H_\text{p}(X|B)]\circledast\mathcal{F}_B[\Pr(B)](0).
\end{align}
The $i$th posterior is used as the $i+1$th prior, therefore we can write the prior resulting from an extended sequence of measurement as:
\begin{multline}
    \mathcal{F}_B\left[\Pr(B|\{X_i=x_i, \tau_i, \theta_i\}_{i=1}^{n})\right](\xi) =\cr\frac{\Conv_{i=1}^n \mathcal{F}_B\left[\Pr(X=x_i|B, \theta_i,\tau_i)\right]\circledast\mathcal{F}_B\left[\Pr(B)\right](\xi)}{\Conv_{i=1}^n \mathcal{F}_B\left[\Pr(X=x_i|B, \theta_i,\tau_i)\right]\circledast\mathcal{F}_B\left[\Pr(B)\right](0)}.
\end{multline}
Each of the terms in the repeated convolution adds three additional peaks in the Fourier domain (with collocation of those peaks being possible). In general, the combination of these peaks is non-trivial, in the case in which all peaks are non-overlapping the number of peaks increase exponentially, in the case that peaks are collocated, the superposition of differing phase coefficients interfere. In the case that the myopic algorithm is followed, the analysis simplifies significantly.

\subsection{Optimisation using Fourier coefficients}

Maximisation of mutual information can be achieved by simultaneously maximising measurement entropy, $H(X)$, and minimising conditional entropy $H(X|B)$. Given the limited form of $\mathcal{F}_B\left[\Pr(B)\right](\xi)$ and $\mathcal{F}_B\left[H_\text{p}(X|B)\right](\xi)$, these quantities can be extracted simply in terms of the prior distribution and measurement parameters ($\theta, \tau$), specifically $H(X)$ is monotonic in the magnitude of the bias and thus maximal values of $H(X)$ are found when $|\Pr(X=x_i)-1/2|$ is minimal:
\begin{align}
    \Pr(X=x_i) &= \mathcal{F}_B[\Pr(X=x_i|B)]\circledast\mathcal{F}_B[\Pr(B)](0)\\
    &= \frac{1}{4}(2\delta(\xi) + e^{-\tau/T-(2\pi  \mu\tau\theta +\pi x_i) i}\delta(\xi+2\mu\tau)\nonumber\\
    &\quad+ (e^{-\tau/T+(2\pi  \mu\tau\theta +\pi x_i)i}\delta(\xi-2\mu\tau))\circledast\mathcal{F}_B[\Pr(B)](0)\\
    &=\frac{2\mathcal{F}_B[\Pr(B)](0)}{4}\nonumber\\
    &\quad+\frac{e^{-\tau/T}(e^{(2\pi  \mu\tau\theta +\pi x_i)i}\mathcal{F}_B[\Pr(B)](2\mu\tau)}{4}\nonumber\\
    &\quad+\frac{e^{-\tau/T}(e^{-(2\pi  \mu\tau\theta +\pi x_i)i}\mathcal{F}_B[\Pr(B)](-2\mu\tau)}{4}
\end{align}
    %\intertext{\textcolor{red}{Bill can you have a look at 56-57?} Since $\Pr(B)$ is normalised, $\mathcal{F}_B[\Pr(B)](0)=1$, hence:}
\begin{align}
    \bigg|\Pr(X=x_i)&-\frac{1}{2}\bigg|\\
    &=e^{-\tau/T}\left|\frac{(e^{(2\pi  \mu\tau\theta +\pi x_i)i}\mathcal{F}_B[\Pr(B)](2\mu\tau)}{4}\right.\nonumber\\
    &\quad+\left.\frac{(e^{-(2\pi  \mu\tau\theta +\pi x_i)i}\mathcal{F}_B[\Pr(B)](-2\mu\tau)}{4}\right|.\label{eq: bias from fourier}
\end{align}
This means that $H(X)$ is minimised in any of the following cases: 
\begin{align}
    \frac{\tau}{T}&\rightarrow\infty\label{eq: no bias decoherence}\\
    \mathcal{F}_B[\Pr(B)](-2\mu\tau)&=\mathcal{F}_B[\Pr(B)](2\mu\tau)=0\label{eq: no bias no frequency component}\\
    \frac{e^{(2\pi  \mu\tau\theta +\pi x_i)i}\mathcal{F}_B[\Pr(B)](2\mu\tau)}{4}&=-\frac{e^{-(2\pi  \mu\tau\theta +\pi x_i)i}\mathcal{F}_B[\Pr(B)](-2\mu\tau)}{4}\label{eq: no bias orthogonal phase}.
\end{align}
These three cases qualitatively represent decoherence removing any bias from the measurement, the prior containing no frequency components that match the measurement, and the phase of the frequency component matching the measurement being orthogonal to the phase of the measurement respectively.

The conditional entropy is more complex, as shown in App.~\ref{app: monotonicity of fourier coefficients}, $\mathcal{F}_B[H_\text{p}(X|B)]$ contains an infinite number of peaks constantly spaced (with a spacing equal to $4\mu\tau$). As shown in App.~\ref{app: monotonicity of fourier coefficients}, in the absence of decoherence ($T\rightarrow\infty$), when $\theta=0$, the coefficients of non-zero frequency peaks are negative, and decreasing in magnitude. Using the Fourier shift theorem, non-zero values of $\theta$ can be included:
\begin{align}
    H(X|B)&=\mathcal{F}_B[H_\text{p}(X|B)]\circledast\mathcal{F}_B[\Pr(B)](0)\\
    &= \left[e^{i\pi\theta\xi}\sum_k \alpha_k \delta(\xi-4\mu\tau k)\right]\circledast\left[\mathcal{F}_B[\Pr(B)](\xi)\right](0)\\
    &=\sum_k e^{4i\pi\theta\mu\tau k}\alpha_k \mathcal{F}_B[\Pr(B)](4\mu\tau k)
    \intertext{The normalisation of $\Pr(B)$ and the unique nature of $\alpha_0$ can be seperated out, simplifying the series $|\alpha_k|$ to a monotonically decreasing real series:}
    H(X|B)&=\alpha_0 - \sum_{k>0} e^{4i\pi\theta\mu\tau k}|\alpha_k| \mathcal{F}_B[\Pr(B)](4\mu\tau k)\label{eq: H(X|B) fourier coeffs form}
    \intertext{The phase components can be absorbed into the effective position of the prior:}
    H(X|B)&=\alpha_0 - \sum_{k>0} |\alpha_k| \mathcal{F}_B\left[\Pr\left(B-\frac{\theta}{2\mu\tau}\right)\right](4\mu\tau k)\label{eq: H(X|B) fourier coeffs form shifted}
\end{align}
%\textcolor{red}{Bill, can you sort out the frequencies within the delta here}
Minimisation of $H(X|B)$ can be achieved my maximising the sum of components of shifted prior over the evenly spaced array defined by $4\mu\tau$, weighted by the monotonically decreasing series $|\alpha_k|$.

\subsection{Inductive behaviour of posterior distribution under myopic optimisation, where $T, \sigma_{\Pr(B)}\rightarrow\infty$}

In the simplified case where decoherence is ignored ($T\rightarrow\infty$) and no prior information is assumed (\emph{i.e.}, the prior is diffuse, $\sigma_{\Pr(B)}\rightarrow\infty$) the analysis of\iffalse Eqs.\fi~\eqref{eq: no bias decoherence}-\eqref{eq: no bias orthogonal phase} and \eqref{eq: H(X|B) fourier coeffs form} are greatly simplified. %\textcolor{red}{Bill, there are some limit concerns here.}

\subsubsection{First measurement case $n=1$}
Optimising the first measurement is trivial. Since the Fourier transform of the diffuse prior contains no frequency information, \emph{i.e.}:
\begin{align}
    \mathcal{F}_B[\Pr(B)](\xi) &=\begin{cases} 
      1 & \xi= 0 \\
      0 & \xi \neq 0 
   \end{cases}\\
   &\eqqcolon \delta^*(\xi)
\end{align}
This implies that for any value of $\tau,\, \theta$ $H(X)$ is maximised and $H(X|B)$ is minimised. To show there is no bias:
\begin{align}
    \Bigg|\Pr(&X=x_i)-\frac{1}{2}\Bigg|\\&=e^{-\tau/T}\left|\frac{(e^{(2\pi  \mu\tau\theta +\pi x_i)i}\times\mathcal{F}_B[\Pr(B)](2\mu\tau)}{4}\right.\nonumber\\
    &\quad+\left.\frac{(e^{-(2\pi  \mu\tau\theta +\pi x_i)i}\times\mathcal{F}_B[\Pr(B)](-2\mu\tau)}{4}\right|\\
    &=e^{0}\left|\frac{e^{(2\pi  \mu\tau\theta +\pi x_i)i}(0)}{4}+\frac{e^{-(2\pi  \mu\tau\theta +\pi x_i)i}(0)}{4}\right|\\
    &= 0.
\end{align}
To show $H(X|B)$ is minimised:
\begin{align}
    H(X|B)&=\alpha_0 - \sum_{k>0} e^{4i\pi\theta\mu\tau k}|\alpha_k| \mathcal{F}_B[\Pr(B)](4\mu\tau k)\\
    &=\alpha_0 - \sum_{k>0} e^{4i\pi\theta\mu\tau k}|\alpha_k|\times(0)\\
    &=\alpha_0.
\end{align}
Since both quantities $H(X)$ and $H(X|B)$ can be simultaneously optimised, and any values of $\tau,\,\theta$ optimise them, any choice of $\tau,\,\theta$ maximise the mutual information $I(X|B)$. Once this measurement has been conducted, the posterior distribution has the following form:
\begin{multline}
    \mathcal{F}_B[\Pr(B|X=x_1)](\xi)= \cr \frac{2\delta^*(\xi) + e^{i\pi(2\theta\mu\tau+x_1)}\delta^*(\xi-2\mu\tau) + e^{-i\pi(2\theta\mu\tau+x_1)}\delta^*(\xi+2\mu\tau)}{4\Pr(X=x_1)}.
\end{multline}
This can be re-written in a shifted parameter:
\begin{multline}
    \mathcal{F}_B\left[\Pr\left(B-\frac{\theta}{2\mu\tau}+\pi x_1\bigg| X=x_1\right)\right](\xi) \cr=\frac{2\delta^*(\xi) + \delta^*(\xi-2\mu\tau) + \delta^*(\xi+2\mu\tau)}{4\Pr(X=x_1)},
\end{multline}

\subsubsection{Inductive case $n\implies n+1$}

The prior that arising from the myopic method after $n$ measurements has the form:%\textcolor{red}{Bill, can you have a look at this definition as well. }
\begin{multline}
    \mathcal{F}_B\left[\Pr\left(B'| \{X_i=x_i\}_{i=1}^n\right)\right](\xi) =\cr\sum_{j=-2^n+1}^{2^n-1} \left(1-\frac{|j|}{2^{n}}\right)\delta^*(\xi-2^{-n+2}\mu\tau_1j).\label{def: general form of prior (myopic)}
\end{multline}
%\textcolor{red}{There's an issue here with the term inside the delta (and that should be a Kroenecker delta)}
From here, we can calculate \iffalse Eqs.~\fi\eqref{eq: bias from fourier} and \eqref{eq: H(X|B) fourier coeffs form} using \iffalse Eq.~\fi\eqref{def: general form of prior (myopic)}:
\begin{align}
    &\left|\Pr(X_{n+1}=x_{n+1})-\frac{1}{2}\right|=\Bigg|\frac{e^{2\pi  \mu\tau_{n+1}\theta_{n+1}i}}{4}\\&\times\left.\left[\sum_{j=-2^n+1}^{2^n-1} \left(1-\frac{|j|}{2^n}\right)\delta^*(2\mu\tau_{n+1}-2^{-n+2}\mu\tau_1j)\right]\right.\nonumber\\
    &+\frac{e^{-2\pi  \mu\tau_{n+1}\theta_{n+1}i}}{4}\nonumber\\
    &\times\left.\left[\sum_{j=-2^n+1}^{2^n-1} \left(1-\frac{|j|}{2^n}\right)\delta^*(-2\mu\tau_{n+1}-2^{-n+2}\mu\tau_1j)\right]\right|,
\end{align}
Looking then at the cases for which bias can be minimised (as shown in \iffalse Eqs.~\fi\eqref{eq: no bias decoherence}-\eqref{eq: no bias orthogonal phase}). In the absence of decoherence, \iffalse Eq.~\fi\eqref{eq: no bias decoherence} cannot be satisfied, however \iffalse Eq.~\fi\eqref{eq: no bias no frequency component} and \eqref{eq: no bias orthogonal phase} can both be satisfied, in the case of \iffalse Eq.~\fi\eqref{eq: no bias no frequency component} this can be achieved by ensuring that: 

\begin{align}
    \lnot\exists k\in \mathbb{N}\,\text{st}:\,\tau_{n+1}=k\frac{\tau_1}{2^{n-1}},\label{eq: optimal tau adaptive 1}
\end{align}
in the case of \iffalse Eq.~\fi\eqref{eq: no bias orthogonal phase} this can be achieved by choosing $\tau_{n+1},\,\theta_{n+1}$:
\begin{align}
    \tau_{n+1} &=k\frac{\tau_1}{2^{n-1}},\,\qquad k\in\mathbb{N}\\
    \theta_{n+1} &= \Delta\theta+\frac{\pi}{2}\\
    \Delta\theta &= 2\mu(B-B')\tau_{n+1}.
\end{align}
Here $\Delta\theta$ acts as a moving ``zero point" that identifies the central peak of the prior distribution.

To maximise mutual information, we must also minimise $H(X|B)$, using \iffalse Eq.~\fi\eqref{eq: H(X|B) fourier coeffs form shifted}:
\begin{multline}
    H(X_{n+1}|B,\{X_i=x_i\}_{i=1}^n) =\\ \alpha_0 - \sum_{k>0} |\alpha_k|e^{i4\pi\mu\tau_{n+1}k\Delta\theta} \\\times\sum_{j=-2^n+1}^{2^n-1} \left(1-\frac{|j|}{2^n}\right)\delta^*(4\mu\tau_{n+1} k-2^{-n+2}\mu\tau_1j).\label{eq: fourier H(X|B) iterative calculation}
\end{multline}
To minimise \iffalse Eq.~\fi\eqref{eq: fourier H(X|B) iterative calculation} it is needed to maximise the sum, this can be achieved by ensuring that the phase term $\text{exp}(i4\pi\mu\tau_{n+1}k\Delta\theta)=1$. This is equivalent to stating $\Delta\theta=0$, reducing \iffalse Eq.~\fi\eqref{eq: fourier H(X|B) iterative calculation} to:
\begin{align}
    &\quad H(X_{n+1}|B,\{X_i=x_i\}_{i=1}^n) \\&= \alpha_0 - \sum_{k>0} |\alpha_k| \sum_{j=-2^n+1}^{2^n-1} \left(1-\frac{|j|}{2^n}\right)\delta^*(4\mu\tau_{n+1} k-2^{-n+2}\mu\tau_1j)\\
    &= \alpha_0 - \sum_{k>0} |\alpha_k| \sum_{j=-2^n+1}^{2^n-1} \left(1-\frac{|j|}{2^n}\right)\delta^*(\tau_{n+1} k-2^{-n}\tau_1j)\\
    &=\alpha_0 - \begin{cases}
        0, & \tau_{n+1} \neq k\frac{\tau_1}{2^{n}}\,\,  \forall k>0\in\mathbb{N}\\
        \sum_{j=-2^n+1}^{2^n-1} |\alpha_{j/m}|\left(1-\frac{|j|}{2^n}\right)&\exists m\in\mathbb{N}\text{ s.t } \tau_{n+1} = m\frac{\tau_1}{2^{n}}
    \end{cases}
\end{align}
Since $|\alpha_k|$ and $\left(1-\frac{|j|}{2^n}\right)$ are both monotonically decreasing functions (for $j,k>0$) the maximum value of their elementwise product occurs when $m=1$. Any other value for $m>1$ drop terms from the sum and reduce the coefficients of the terms included. Under these conditions the minimal conditional entropy can be written
\begin{align}
     H(X_{n+1}|B,\{X_i=x_i\}_{i=1}^n) &= \alpha_0 - \sum_{j=1}^{2^n-1} |\alpha_k|\left(1-\frac{|j|}{2^n}\right)\\
     \iff \tau_{n+1} &= \frac{1}{2^{n-1}}\tau_1,\,\theta_{n+1}=\Delta\theta.\label{eq: optimal tau theta 1}
\end{align}
Notably, this restriction on $\tau_{n+1}, \theta_{n+1}$ is compatible with \iffalse Eq.~\fi\eqref{eq: optimal tau adaptive 1} allowing for simultaneous optimisation of $H(X_{n+1}|\{X_i=x_i\}_{i=1}^n)$ and $H(X_{n+1}|B,\{X_i=x_i\}_{i=1}^n)$, in turn making it an optimal solution for maximising $I(X:B|\{X_i=x_i\}_{i=1}^n)$.

Finally, to complete the proof, we must show that the form of the posterior is consistent with \iffalse Eq.~\fi\eqref{def: general form of prior (myopic)}, specifically:

\begin{multline}
    \mathcal{F}_B\left[\Pr\left(B''| \{X_i=x_i\}_{i=1}^{n+1}\right)\right](\xi) =\\
    \frac{\mathcal{F}_B\left[\Pr\left(B'| \{X_i=x_i\}_{i=1}^n\right)\right]\circledast\mathcal{F}_B\left[\Pr\left(
    X_{n+1}=x_{n+1}|B',\tau_{n+1},\theta_{n+1}\right)\right](\xi)}{\Pr(X_{n+1}=x_{n+1})}
\end{multline}
Since $X_{n+1}$ maximises measurement entropy, it is unbiased, and thus $\Pr(X_{n+1}=x_{n+1}) =1/2$, hence:
\begin{align}
    \mathcal{F}_B&\left[\Pr\left(B''| \{X_i=x_i\}_{i=1}^{n+1}\right)\right](\xi)\\
    &=2\sum_{j=-2^n+1}^{2^n-1} \left(1-\frac{|j|}{2^n}\right)\delta^*(\xi-2^{-n+2}\mu\tau_1j)\nonumber\\
    &\quad\circledast\frac{2\delta^*(\xi) + e^{i\pi x_{n+1}}\delta^*(\xi-2\mu\tau_{n+1}) + e^{-i\pi x_{n+1}}\delta^*(\xi+2\mu\tau_{n+1})}{4}\\
    &=\sum_{j=-2^n+1}^{2^n-1} \left(1-\frac{|j|}{2^n}\right)\delta^*(\xi-2^{-n+2}\mu\tau_1j)\nonumber\\
    &\quad+\frac{e^{i\pi x_{n+1}}}{2}\sum_{j=-2^n+1}^{2^n-1} \left(1-\frac{|j|}{2^n}\right)\delta^*(\xi-2^{-n+2}\mu\tau_1j-2\mu\tau_{n+1})\nonumber\\
    &\quad+\frac{e^{-i\pi x_{n+1}}}{2}\sum_{j=-2^n+1}^{2^n-1} \left(1-\frac{|j|}{2^n}\right)\delta^*(\xi-2^{-n+2}\mu\tau_1j+2\mu\tau_{n+1})
\end{align}
Since $\tau_{n+1}=\tau_12^{-n+1}$:
\begin{align}
    \mathcal{F}_B&\left[\Pr\left(B''| \{X_i=x_i\}_{i=1}^{n+1}\right)\right](\xi)\\&=\sum_{j=-2^n+1}^{2^n-1} \left(1-\frac{|j|}{2^n}\right)\delta^*(\xi-2^{-n+2}\mu\tau_1j)\nonumber\\
    &\quad+\frac{e^{i\pi x_{n+1}}}{2}\sum_{j=-2^n+1}^{2^n-1} \left(1-\frac{|j|}{2^n}\right)\delta^*(\xi-2^{-n+2}\mu\tau_1j-2^{-n+1}\mu\tau_{1})\nonumber\\
    &\quad+\frac{e^{-i\pi x_{n+1}}}{2}\sum_{j=-2^n+1}^{2^n-1} \left(1-\frac{|j|}{2^n}\right)\delta^*(\xi-2^{-n+2}\mu\tau_1j+2^{-n+1}\mu\tau_1)\\
    &=\sum_{j=-2^n+1}^{2^n-1} \left(1-\frac{|j|}{2^n}\right)\delta^*(\xi-2\cdot2^{-n+1}\mu\tau_1j)\nonumber\\
    &\quad+\frac{e^{i\pi x_{n+1}}}{2}\sum_{j=-2^n+1}^{2^n-1} \left(1-\frac{|j|}{2^n}\right)\delta^*(\xi-2^{-n+1}\cdot(2j-1)\mu\tau_{1})\nonumber\\
    &\quad+\frac{e^{-i\pi x_{n+1}}}{2}\sum_{j=-2^n+1}^{2^n-1} \left(1-\frac{|j|}{2^n}\right)\delta^*(\xi-2^{-n+1}\cdot(2j+1)\mu\tau_{1})
\end{align}
Here we can see separation into an even and two odd arrays, offset by 2. Refactoring into $j'=2j$ or $j'=2j\pm1$:
\begin{multline}
    \mathcal{F}_B\left[\Pr\left(B''| \{X_i=x_i\}_{i=1}^{n+1}\right)\right](\xi)=\\\sum_{\substack{j'=-2^{n+1}+1\\ j'\in\text{Even}}}^{2^{n+1}-1} \left(1-\frac{|j'|}{2^{n+1}}\right)\delta^*(\xi-2^{-n+1}\mu\tau_1j')\nonumber\\
    \quad+e^{i\pi x_{n+1}}\sum_{\substack{j'=-2^{n+1}+1\\ j'\in\text{Odd}}}^{2^{n+1}-1} \left(1-\frac{|j'|}{2^{n+1}}\right)\delta^*(\xi-2^{-n+1}\mu\tau_{1}j')
\end{multline}
The phase component can be absorbed into $B''$ if we use the adaption $B''=B'+\frac{\pi x_{n+1}}{2\mu\tau_{n+1}}$, allowing the sums to be combined, yeilding the desired form:
\begin{multline}
    \mathcal{F}_B\left[\Pr\left(B''| \{X_i=x_i\}_{i=1}^{n+1}\right)\right](\xi)\\
    =\sum_{j'=-2^{n+1}+1}^{2^{n+1}-1} \left(1-\frac{|j'|}{2^{n+1}}\right)\delta^*(\xi-2^{-n+1}\mu\tau_1j').
\end{multline}
This derivation also implicitly solves the value of $\Delta\theta$. We can track $B'$ iteratively:
\begin{align}
    B'_{n+1} &= B'_{n} + \frac{\pi x_{n+1}}{}\\
    \implies \Delta\theta_{n+1}&=\Delta\theta_n\frac{\tau_{n+1}}{\tau_{n}} + \frac{\pi x_{n+1}\tau_{n+1}}{\tau_{n}}\\
    &=\frac{\Delta\theta_n+\pi x_{n+1}}{2}
\end{align}
Since the optimal choice of $\theta_n$ exactly coincides with $\Delta\theta_n$, we can write the completed iterative expressions for $\tau_n,\,\theta_n$:
\begin{align}
    \tau_{n+1} &= \frac{1}{2}\tau_n\\
    \theta_{n+1} &= \frac{\theta_n+\pi x_{n}}{2}.
\end{align}
This can also be written explicitly:
\begin{align}
    \tau_{n+1} &= \frac{1}{2^n}\tau_0\\
    \theta_{n+1} &= \frac{1}{2^n}\theta_0+\sum_{i=1}^n\frac{\pi x_{i}}{2^{n-i}},
\end{align}
where $\tau_0, \theta_0$ are two free parameters that can be chosen without effecting the entropy extraction rate (in simplified case).

%\section{Analysis of Fourier coefficients}
\section{Analysis of $\mathcal{F}[H_\text{p}(X|B)]$}
\label{app: monotonicity of fourier coefficients}
To examine the properties of the Fourier transform of $H_P$, in the case of the likelihoods from \iffalse Eqs.~\fi\eqref{def: pr(x=0) likelihood function} and \eqref{def: pr(x=1) likelihood function} we directly calculate the following integral.
\begin{multline}
	\label{eq:adad}
	\alpha_{j}=\frac{4\mu\tau}{\pi}\int_{-\frac{\pi}{4\mu\tau}}^{\frac{\pi}{4\mu\tau}} \left[-\left(\frac{1+\cos\left(2\mu B\tau +\theta\right)}{2}\right)\right.\\
    \ln\left(\frac{1+\cos(2\mu B\tau +\theta)}{2}\right)-\left.\left(\frac{1-\cos(2\mu B\tau +\theta)}{2}\right)\right.\\
    \left.\ln\left(\frac{1-\cos(2\mu B\tau +\theta)}{2}\right)\right]\left(\cos(2j(2\mu B\tau +\theta))\right)dB
\end{multline}

A simple change of variable $2\mu \tau dB \to dx$ gives:
\begin{multline}
	\label{eq:2}
	\text{Int}_{j} = -\frac{2}{\pi}\int_{-\frac{\pi}{2}+\theta}^{\frac{\pi}{2}+\theta} \left[\left(\frac{1+\cos(x)}{2}\right)\ln\left(\frac{1+\cos(x)}{2}\right)\right.\\ +\left.\left(\frac{1-\cos(x)}{2}\right)\ln\left(\frac{1-\cos(x)}{2}\right)\right]\left(\cos(2jx)\right)dx
\end{multline}
It is easy to see, using $\cos(x+\pi)=-\cos(x)$ and periodicity,  that this reduces to
\begin{equation}
	\label{eq:4}
	\text{Int}_{j} = -\frac{2}{\pi}\int_{0}^{2\pi} \left(\frac{1+\cos(x)}{2}\right)\ln\left(\frac{1+\cos(x)}{2}\right)\cos(2jx)\,dx;
\end{equation}
the (cosine) Fourier coefficients of $\left(\frac{1+\cos(x)}{2}\right)\ln\left(\frac{1+\cos(x)}{2}\right)$.
Write $z=e^{ix}$, so that \eqref{eq:4} becomes
\begin{multline}
    \label{eq:4222}
	\text{Int}_{j} = -\frac{1}{2\pi}\int_{0}^{2\pi} \left(1+\frac12\left(z+\frac1z\right)\right)\cr\ln\left(\frac{1+\frac12\left(z+\frac1z\right)}{2}\right)\left(z^{2j}+z^{{-2j}}\right)\,dx;
\end{multline}
We write, for $0<\epsilon< \frac12$,
\begin{multline*}
	\label{eq:4222ddd}
	\text{Int}_{j}(\epsilon) = -\frac{1}{2\pi}\int_{0}^{2\pi} \left(1+\frac1{2+\epsilon}\left(z+\frac1z\right)\right)\cr\ln\left(\frac{1+\frac1{(2+\epsilon)}\left(z+\frac1z\right)}{2}\right)\left(z^{2j}+z^{{-2j}}\right)\,dx.
\end{multline*}
This modification enables Laurent expansions in an annulus of the form
\begin{equation*}
	A_{{r,R}}=\{z: r<|z|<R\},
\end{equation*}
where $r<1<R$. For suitable $r$ and $R$, if $z\in A_{r,R}$, then
\begin{equation*}
	|u|=\left|z+\frac1z\right| < 2+\epsilon,
\end{equation*}
and so the power series expansion of $\log(1+\frac1{2+\epsilon}u)$ converges in that annulus.
Moreover, since the integrand is bounded in absolute value for all $\epsilon$, by the Lebesgue Dominated Convergence Theorem,
$\text{Int}_{j}(\epsilon)\to \text{Int}_{j} \text{ as } \epsilon\to 0$.

Symmetry of the integrand with respect to $z\to\frac1z$ gives
\begin{multline}
    \label{eq:4222dddddd}
	\text{Int}_{j}(\epsilon) = -\frac{1}{\pi}\int_{0}^{2\pi} \left(1+\frac1{2+\epsilon}\left(z+\frac1z\right)\right)\cr\ln\left(\frac{1+\frac1{(2+\epsilon)}\left(z+\frac1z\right)}{2}\right)z^{2j}\,dx.
\end{multline}
Moving to a contour integral, we obtain
\begin{multline}
	\label{eq:4rrr}
	\text{Int}_{j}(\epsilon) = -\frac{1}{\pi i}\int_{\Gamma} \left(1+\frac1{2+\epsilon}\left(z+\frac1z\right)\right)\cr\ln\left(\frac{1+\frac1{(2+\epsilon)}\left(z+\frac1z\right)}{2}\right)z^{2j}\,\frac{dz}z,
\end{multline}
where $\Gamma$ is the unit circle.
Now, we define
\begin{equation}
	\label{eq:3}
	L_{k}(\epsilon) = -\frac{1}{\pi i}\int_{\Gamma}\ln\left(\frac{1+\frac1{(2+\epsilon)}\left(z+\frac1z\right)}{2}\right)z^{k}\,\frac{dz}z,
\end{equation}
so that,
\begin{equation}
	\label{eq:5}
	\text{Int}_{j}(\epsilon) =L_{2j}(\epsilon) +\frac1{2+\epsilon}(L_{2j-1}(\epsilon)+L_{2j+1}(\epsilon)).
\end{equation}
We also write $L_{k}$ for $L_{k}(0)$.

Observe that, since we are only interested in $k\ge 1$,
%\begin{equation}
%	\label{eq:30}
%	\frac{1}{\pi i}\int_{\Gamma}\log(2)z^{k}\,\frac{dz}z = 0,
%\end{equation}
%so that
\begin{equation}
	\label{eq:3ddd}
	L_{k}(\epsilon) = -\frac{1}{\pi i}\int_{\Gamma}\ln\left(1+\frac1{(2+\epsilon)}{\left(z+\frac1z\right)}\right)z^{k}\,\frac{dz}z,
\end{equation}
By the Calculus of Residues for an annulus,
\begin{equation}
	\label{eq:1}
	L_{k}(\epsilon)=-2\text{Coeff}_{0}\left[\ln\left(1+\frac1{(2+\epsilon)}{\left(z+\frac1z\right)}\right)z^{k}\right],
\end{equation}
where
$\text{Coeff}_{0}(f(z))$ is the constant term in the Laurent expansion of $f(z)$ at $0$.

In the annulus $A_{r,R}$,
\begin{equation}
	\label{eq:31}
	\ln\left(1+\frac1{(2+\epsilon)}{\left(z+\frac1z\right)}\right)=\sum_{n=1}^{\infty} \frac{(-1)^{n-1}}{n} \frac1{(2+\epsilon)^{n}}\left(z+\frac1z\right)^{n},
\end{equation}
so that
\begin{multline*}	\text{Coeff}_{0}\left[\ln\left(1+\frac1{(2+\epsilon)}{\left(z+\frac1z\right)}\right)z^{k}                      \right]\\=\sum_{n=1}^{\infty}\frac{(-1)^{n-1}}{n} \frac1{(2+\epsilon)^{n}}\text{Coeff}_{0}\left[\left(z+\frac1z\right)^{n}z^{k}\right].
\end{multline*}
It is clear that $\text{Coeff}_{0}\left[\left(z+\frac1z\right)^{n}z^{k}\right]$ is non-zero only if $n\ge k$ and $n-k$ is even. We look at the $k$ odd and $k$ even cases separately. For $k=2\ell$ even,
\begin{equation*}
	\text{Coeff}_{0}\left[\left(z+\frac1z\right)^{2m}z^{2\ell}\right] = \binom{2m}{m+\ell},
\end{equation*}
yielding
\begin{equation}
	\label{eq:6}
	L_{2\ell}(\epsilon)=2\sum_{m=\ell}^{\infty} \frac1{2m}\frac1{(2+\epsilon)^{2m}}\binom{2m}{m+\ell}.
\end{equation}
Similarly, for $k=2\ell+1$, odd,
\begin{equation*}
	\text{Coeff}_{0}\left[\left(z+\frac1z\right)^{2m+1}z^{2\ell+1}\right] = \binom{2m+1}{m+\ell+1},
\end{equation*}
and this gives,
\begin{equation}
	\label{eq:6a}
	L_{2\ell+1}(\epsilon)=-2\sum_{m=\ell}^{\infty} \frac{1}{2m+1} \frac1{(2+\epsilon)^{2m+1}} \binom{2m+1}{m+\ell+1}.
\end{equation}

It is safe, at this stage, to let $\epsilon\to0 $ by, say, the Lebesgue Monotone Convergence Theorem, so that
\begin{align*}
    L_{2\ell}    &=2\sum_{m=\ell}^{\infty} \frac1{2m}\frac1{2^{2m}}\binom{2m}{m+\ell}\\
    L_{2\ell+1}  &=-2\sum_{m=\ell}^{\infty} \frac{1}{2m+1} \frac1{2^{2m+1}} \binom{2m+1}{m+\ell+1}.
\end{align*}

Now, from \eqref{eq:5},
% \begin{multline*}
% 	\label{eq:7}
% 	\text{Int}_{j}(\epsilon) = -\sum_{m=j}^{\infty} \frac1{2m}\frac1{2^{2m}}\binom{2m}{m+j}+
% 	\sum_{m=j}^{\infty} \frac{1}{2m-1} \frac1{2^{2m-1}} \binom{2m-1}{m+j-1}\\ +
% 	\sum_{m=j}^{\infty} \frac{1}{2m+1} \frac1{2^{2m+1}} \binom{2m+1}{m+j+1}.
% \end{multline*}
\begin{multline*}
	%	\label{eq:7}
	\text{Int}_{j} = \sum_{m=j}^{\infty}\frac1{2^{2m}}\biggl[ \frac1{m}\binom{2m}{m+j}
		-	\frac{4}{2m-1}  \binom{2m-1}{m+j-1}\\-
		\frac{1}{2m+1}  \binom{2m+1}{m+j+1}\biggr].
\end{multline*}
We write $\alpha(m,j)$ for the term in square brackets:
\begin{equation*}
	\alpha(m,j)= 2\binom{2m-1}{m+j-1}\left(\frac{1}{m+j}\\-\frac2{2m-1}-\frac{m}{(m+j+1)(m+j)}   \right).
\end{equation*}
Straightforward calculations show, for $m\ge j\ge 1$,  firstly, that
\begin{equation*}
	\alpha(m,j)=-\frac{1}{m(2m+1)(2m-1)}\binom{2m+1}{m+j+1}\left(2m^{2}+2mj+3j+1
	\right),
\end{equation*}
so  negative and, secondly,  that
\begin{equation*}
	\frac{\alpha(m,j)}{\alpha(m,j+1)}=\frac{4(j + 1)m^2 + 2m^3 + 3j^2 + (2j^2 + 7j + 1)m + 7j + 2}{2m^3 - 3j^2 - (2j^2 - j - 4)m + 2m^2 - 4j},
\end{equation*}
which is always greater than $1$ in the range of values of interest.
It follows that
$\text{Int}_{j}$ is negative and increasing in $j$, as required.

\bibliography{references}
\bibliographystyle{IEEEtran}

%\end{thebibliography}

\begin{IEEEbiography}[{\includegraphics[width=1in,height=1.25in,clip,keepaspectratio]{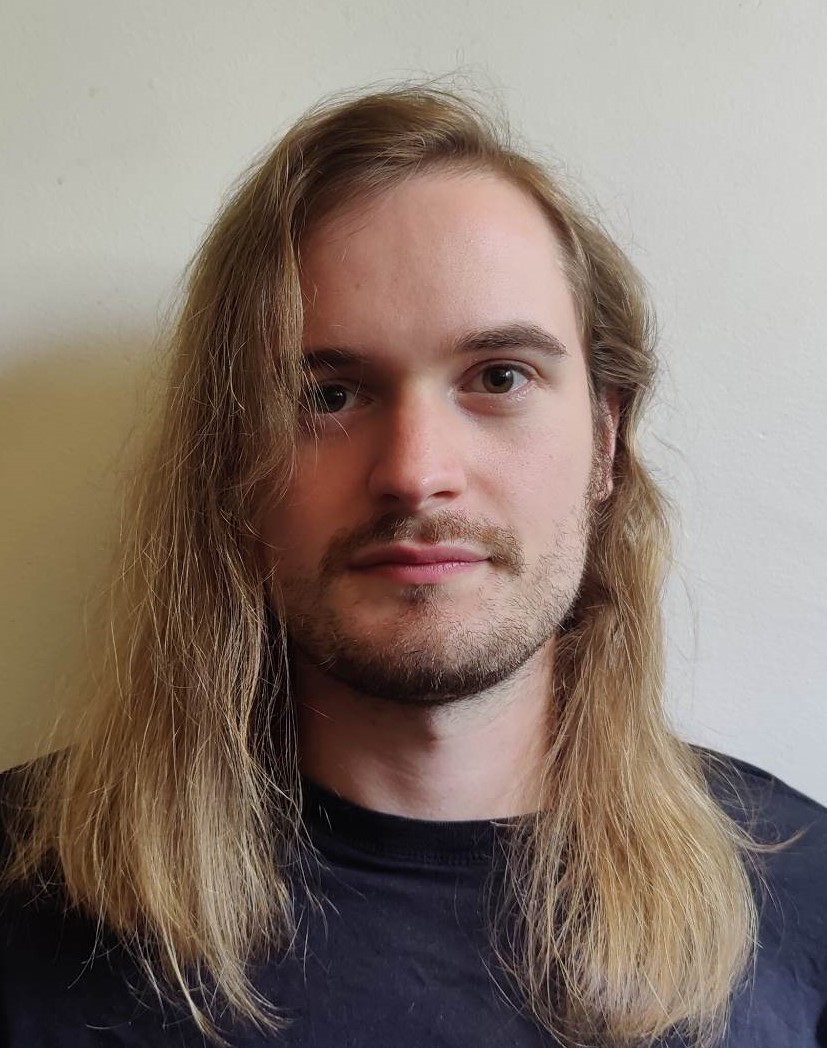}}]{Julian Greentree} received the B.Sci degree from the University of Melbourne in 2019 and B.Sci (Hons) degree in Physics from RMIT University, Melbourne in 2021. He is currently completing a PhD in Quantum information theory at the University of Melbourne.

His research interests include the intersection between classical and quantum information theory, optimal measurement and sensing using quantum systems, and fundamental quantum limits.

Mr. Greentree is currently supported by an RTP grant from the Australian Government
\end{IEEEbiography}

\begin{IEEEbiography}
[{\includegraphics[width=1in,height=1.25in,clip,keepaspectratio]{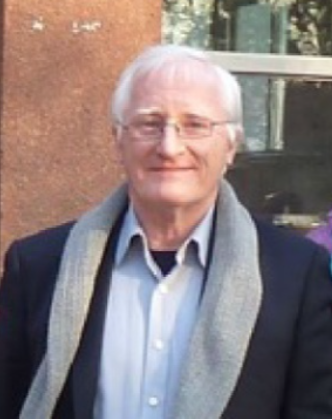}}]{William Moran}
Bill Moran (M’95) is an honorary Professor of Electrical Engineering at the University of Melbourne. Previously, from  2017 to 2022, he was the Professor of Defence Technology in the University of Melbourne. From 2014 to 2017, he was Director of the Signal Processing and Sensor Control Group in the School of Engineering at RMIT University, from 2001 to 2014, a Professor in the Department of Electrical Engineering, University of Melbourne, Director of Defence Science Institute in University of Melbourne (2011-14), Professor of Mathematics (1976–1991), Head of the Department of Pure Mathematics (1977–79, 1984–86), Dean of Mathematical and Computer Sciences (1981, 1982, 1989) at the University of Adelaide, and Head of the Mathematics Discipline at the Flinders University of South Australia (1991–95). He was Head of the Medical Signal Processing Program (1995–99) in the Cooperative Research Centre for Sensor Signal and information Processing. He was a member of the Australian Research Council College of Experts from 2007 to 2009. He was elected to the Fellowship of the Australian Academy of Science in 1984. He holds a Ph.D. in Pure Mathematics from the University of Sheffield, UK (1968), and a First Class Honours B.Sc. in Mathematics from the University of Birmingham (1965). He has been a Principal Investigator on numerous research grants and contracts, in areas spanning pure mathematics to radar development, from both Australian and US Research Funding Agencies, including DARPA, AFOSR, AFRL, Australian Research Council (ARC), Australian Department of Education, Science and Training, and Defence Science and Technology, Australia. His main areas of research interest are in signal processing both theoretically and in applications to radar, waveform design, antenna  and radar theory, optimization and control, sensor networks, and sensor management. He has also worked in various areas of mathematics including harmonic analysis, representation theory, and number theory.
\end{IEEEbiography}

\begin{IEEEbiography}[{\includegraphics[width=1in,height=1.25in,clip,keepaspectratio]{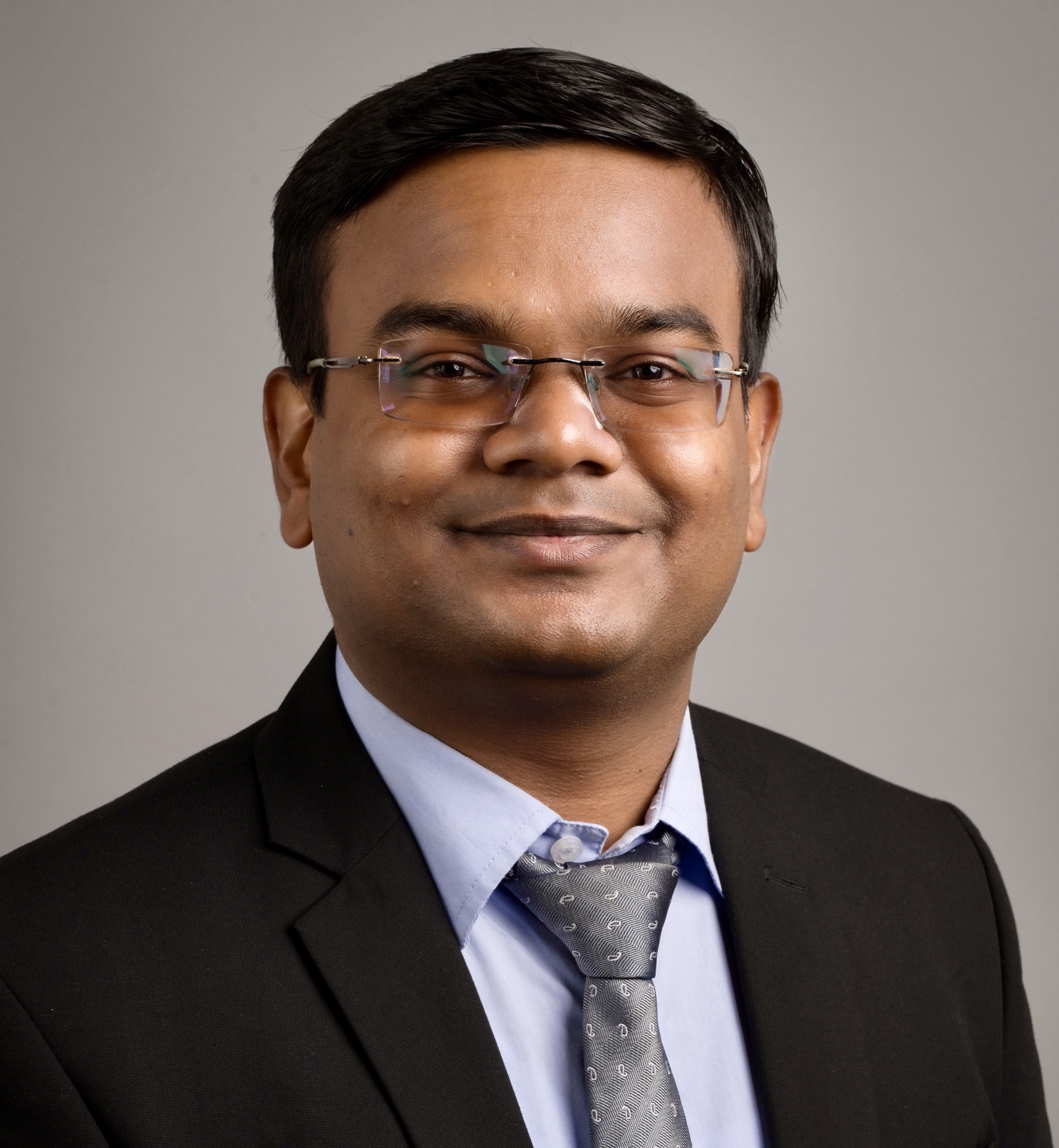}}]
{\bf Neel Kanth Kundu} (Member, IEEE) received the B.Tech. degree in
electrical engineering with a specialization in communication systems
and networking from the Indian Institute of Technology Delhi, New Delhi,
India, in 2018, and the Ph.D. degree in electronic and computer
engineering (ECE) with a concentration in scientific computation from
The Hong Kong University of Science and Technology (HKUST), Clear Water
Bay, Hong Kong in 2022. Since Oct’23, he has been with the Centre for
Applied Research in Electronics (CARE), Indian Institute of Technology
Delhi, as an Assistant Professor.

 From Feb’23 to Oct’23 he was a postdoctoral research fellow with the
Department of Electrical and Electronic Engineering at the University of
Melbourne, Australia and from Sep’23 to Jan’24 he was a postdoctoral
research associate with the department of ECE, HKUST in Hong Kong. From
December 2021 to May 2022, he was a visiting Ph.D. student with the
Department of Engineering, University of Ferrara, Ferrara, Italy. From
May 2016 to July 2016, he was a visiting student intern with the Rice
Integrated Systems and Circuits Lab, Rice University, Houston, TX, USA. 
His research interests include signal processing for 6G wireless
communications, quantum communications, and quantum information
processing.

Dr. Kundu is a recipient of the INSPIRE Young Faculty Fellowship awarded
by the Department of Science and Technology, Government of India. Dr.
Kundu was the recipient of the Hong Kong Ph.D. Fellowship and the
Overseas Research Award at HKUST.
\end{IEEEbiography}

\newpage

%If you do not have or do not want to include a photo, you can use IEEEbiographynophoto as shown below:

\iffalse\begin{IEEEbiographynophoto}{Third C. Author, Jr.} (M'87) received the B.S. degree in mechanical
engineering from National Chung Cheng University, Chiayi, Taiwan, in 2004
and the M.S. degree in mechanical engineering from National Tsing Hua
University, Hsinchu, Taiwan, in 2006. He is currently pursuing the Ph.D.
degree in mechanical engineering at Texas A{\&}M University, College
Station, TX, USA.

From 2008 to 2009, he was a Research Assistant with the Institute of
Physics, Academia Sinica, Tapei, Taiwan. His research interest includes the
development of surface processing and biological/medical treatment
techniques using nonthermal atmospheric pressure plasmas, fundamental study
of plasma sources, and fabrication of micro- or nanostructured surfaces.

Mr. Author's awards and honors include the Frew Fellowship (Australian
Academy of Science), the I. I. Rabi Prize (APS), the European Frequency and
Time Forum Award, the Carl Zeiss Research Award, the William F. Meggers
Award and the Adolph Lomb Medal (OSA).
\end{IEEEbiographynophoto}
\fi

\end{document}